\documentclass[authoryear]{elsarticle}
\usepackage{natbib}
\usepackage{longtable}
\usepackage{lscape}

\usepackage{graphicx}
\begin{document}

\begin{frontmatter}

\title{Making the Earth: Combining Dynamics and Chemistry in the Solar System}

\author[LPL]{Jade C. Bond\corref{cor1}}
%\ead{jbond@lpl.arizona.edu}
\author[LPL]{Dante S. Lauretta}
%\ead{lauretta@lpl.arizona.edu}
\author[PSI]{David P. O'Brien}
%\ead{obrien@psi.edu}

\address[LPL]{LPL, University of Arizona, 1629 E. University Blvd, Tucson AZ 85721, USA}
\address[PSI]{Planetary Science Institute, 1700 E. Fort Lowell, Suite 106, Tucson AZ 85719, USA}
\cortext[cor1]{Corresponding author}

\begin{abstract}
No terrestrial planet formation simulation completed to date has considered the detailed chemical composition of the planets produced. While
many have considered possible water contents and late veneer compositions, none have examined the bulk elemental abundances of the planets
produced as an important check of formation models. Here we report on the first study of this type. Bulk elemental abundances based on disk
equilibrium studies have been determined for the simulated terrestrial planets of \cite{dave}. These abundances are in excellent agreement with
observed planetary values, indicating that the models of \cite{dave} are successfully producing planets comparable to those of the Solar System
in terms of both their dynamical \emph{and} chemical properties. Significant amounts of water are accreted in the present simulations, implying
that the terrestrial planets form ``wet'' and do not need significant water delivery from other sources. Under the assumption of equilibrium
controlled chemistry, the biogenic species N and C still need to be delivered to the Earth as they are not accreted in significant proportions
during the formation process. Negligible solar photospheric pollution is produced by the planetary formation process. Assuming similar levels of
pollution in other planetary systems, this in turn implies that the high metallicity trend observed in extrasolar planetary systems is in fact
primordial.
\end{abstract}

\begin{keyword}
Terrestrial Planets \sep Planetary Formation  \sep Origin, Solar System
%\texttt{http://icarus.cornell.edu/information/keywords.html}
\end{keyword}

\end{frontmatter}

\section{Introduction}
Terrestrial planet formation, both in terms of the dynamics and chemistry involved, is still not fully understood. Dynamically, basic planetary
formation is described through the planetesimal theory (see \cite{chambers} for a detailed review). This theory sees planetary formation
occurring through three main steps. Initially, dust settles into the midplane and accretes to form planetesimals, the first solid bodies of the
system. This stage is followed by the collisional accretion of planetesimals to produce planetary embryos. The growth of the embryos occurs
initially via runaway growth (where an increasing geometric cross-section and gravitational field allow for the accretion of an ever increasing
number of planetesimals) before transitioning to oligarchic growth (where neighboring embryos grow at similar rates). Finally, as numbers of
planetesimals decrease, the interaction between embryos becomes the dominant factor as they perturb each other onto crossing orbits, thus
producing accretion via violent collisions.

Many attempts have been made to simulate the third stage of terrestrial planet formation described above (e.g.
\citealt{kom1,kom2,chambers2,chambers3,raymond04}). However, these simulations have had limited success and no simulation has been able to
\emph{exactly} reproduce the terrestrial planets of the Solar System in terms of their number, masses and orbital parameters. For example,
direct N-body simulations have produced systems with approximately the correct number of planets yet with an orbital excitation greater than
that observed for the Solar System \citep{chambers2,chambers3,raymond04}. Similarly, simulations incorporating tidal torques (e.g.
\citealt{kom1,kom2}) have produced too many planets but with more favorable excitation levels.

Recent developments in modeling have occurred with the incorporation of dynamical friction, the process whereby equipartitioning of energy
between low mass planetesimals and larger embryos results in reduced relative velocities for the embryos, thus increasing their probability of
accreting. Dynamical friction has been shown to be a viable mechanism to reduce the dynamical excitation levels of the final planets to better
agree with observed values within the Solar System \citep{levison}.  Several authors have recently performed N-body simulations with a high
enough resolution to accurately model the effects of dynamical friction during terrestrial planet accretion, such as \citet{dave} and
\citet{raymond:2006,raymond:2007,raymond:2009}.  The simulations of \citet{dave}, which will be used in this work, represent a factor of $\sim$5
increase in the number of gravitationally interacting bodies compared to most other previous simulations of this type (e.g.
\citealt{chambers3}), and because of the large number of small planetesimals in the simulation, dynamical friction is more accurately treated
than in those previous simulations. The terrestrial planets produced in the \citeauthor{dave} simulations are a significantly better fit to the
observed properties of the terrestrial planets of the Solar System than those of previous simulations. There are fewer planets (averaging 3.25
planets per simulation) and the dynamical excitation of the planetary systems is comparable to that of the actual terrestrial planets in the
Solar System. Furthermore, the accretion timescales of the planets simulated are in agreement with the accretion timescale for the Earth
($\sim$60 Myr) as obtained from $^{182}$Hf-$^{182}$W dating \citep{touboul}.  The simulations of \citet{raymond:2006,raymond:2007,raymond:2009},
which have a comparable resolution to the \citeauthor{dave} simulations, also produce a comparable improvement over previous simulations.  Thus
it can be seen that these recent N-body simulations have provided us with some of the most realistic and feasible models of terrestrial planet
formation completed to date, and clearly represent a significant step towards understanding the last stage of the planetary formation.

Several other types of terrestrial planet formation models have been developed recently as well. The effects of the depleting Solar nebula gas
on terrestrial planet accretion, both in terms of its effects through gravitational tides and also by causing the sweeping of secular resonances
through the terrestrial planet region, has been studied analytically by \citet{nagasawa:2005}.  More recently, \citet{thommes:2008} incorporated
these effects into an N-body integrator.  That work did not incorporate a planetesimal swarm (and hence did not include dynamical friction), but
found that under certain circumstances, the masses and dynamical state of the terrestrial planets in the Solar System could be reproduced.
\citet{bromley:2006} and \citet{kenyon:2006} developed a hybrid code in which large bodies are treated via an N-body integration scheme, while
small bodies are treated with a multi-annulus coagulation model that can treat the growth of bodies starting from km-scale planetesimals.  They
too found that their model could reproduce many of the dynamical characteristics of the terrestrial planets in the Solar System, and while the
final eccentricities of the planets were somewhat high, it is likely that a more complete treatment of the effects of fragmentation, and of
tidal drag on large bodies from the nebular gas, would help to decrease them.  A full comparison of these models with the direct N-body
simulations of \citet{dave} and others is beyond the scope of this article, but we will discuss the implications of those models for the work
presented here in Sec.~\ref{sec:discussion}.

While models have advanced significantly in their ability to reproduce the terrestrial planets from a dynamical point of view, the vast majority
of terrestrial planet formation studies completed to date have been limited (at best) in their simultaneous chemical composition studies.
Several studies have estimated the amount of potentially hydrated material that may have accreted into the terrestrial planets (e.g.
\citealt{raymond04,dave,raymond:2007}) while \citet{dave} and \citet{raymond:2007} also considered the delivery of siderophile-element-rich late
veneer material needed to account for the highly siderophile element budget of the Earth's mantle \citep{chou}. However, detailed examination of
the bulk chemical composition of the resulting planets as a test of dynamical simulations has never been thoroughly explored. This combination
of the two approaches to produce a comprehensive model of both the chemistry and dynamics of terrestrial planet formation is essential to
determine how well current numerical simulations reproduce not only the masses and dynamic state of the terrestrial planets, but also their bulk
composition. The present study represents a first step towards such a model.

We have derived detailed bulk elemental abundances for all of the terrestrial planets formed in the simulations of \cite{dave}. These predicted
compositions are compared to the bulk compositions of the actual terrestrial planets as an examination of the chemical plausibility of the
dynamical simulations currently being used to model terrestrial planet formation. This approach allows us to not only examine the bulk elemental
composition of the final planets but to also study the compositional evolution with time. Additionally, we also investigate the delivery of
hydrated material, which has important implications for the development of habitable terrestrial planets, along with the composition of the
``late veneer'' material accreted by the planets. Finally, the amount of material accreted by the Sun as ``pollution'' during the terrestrial
planet formation process and the resulting changes in stellar photospheric abundances are also examined. This study is the first to attempt to
produce a comprehensive, simultaneous model of both the chemistry and dynamics of terrestrial planet formation. As such, it represents a
significant step forward in producing a cohesive model of terrestrial planet formation.

\section{Simulations}
\subsection{Dynamical}
\begin{figure}
\begin{center}
\includegraphics[width=120mm]{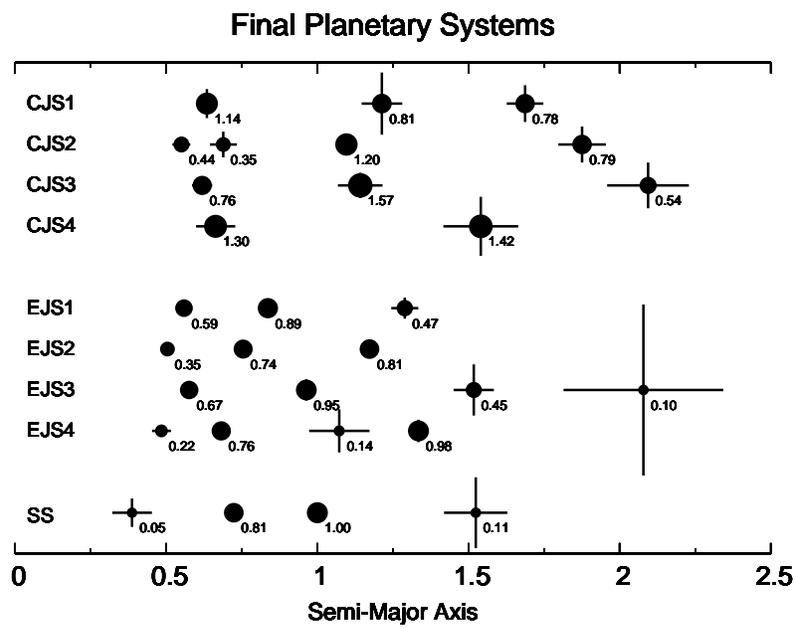} \caption[Schematic of the results of the simulations of \cite{dave}.]
{Schematic of the results of the simulations of \cite{dave}. The horizontal lines indicate the variation between aphelion and
perihelion. The vertical lines indicate variation in distance from the midplane due to the inclination of the planet. Numerical values represent
the mass of the planet in Earth masses. The Solar System (SS) is shown for comparison.\label{final_sys}}
\end{center}
\end{figure}

The eight high-resolution N-body accretion simulations of \citet{dave} were utilized in this study.  These simulations were performed with the
SyMBA integrator \citep{duncan}, which is ideally suited for this type of simulation as it is designed to rapidly and accurately integrate the
orbits of a system of bodies on Keplerian orbits, and can treat close encounters and mergers between colliding bodies. Each simulation consists
of an equal distribution of mass between Mars-mass embryos (0.0933M$_{\bigoplus}$, 25 per simulation) and planetesimals 1/40th the size of the
embryos (0.00233M$_{\bigoplus}$, $\sim$1000 per simulation).  As is common in this type of simulation, the planetesimals gravitationally
interact with the embryos, but not one-another.  The initial distribution of bodies is constrained by the disk surface density profile
$\sum$(r)=$\sum_{0}$(r/1AU)$^{-3/2}$ with $\sum_{0}$ = 8 gcm$^{-2}$ \citep{chambers3} and bodies initially located between 0.3 and 4.0 AU
(\citealt{dave}).   Four simulations were run with Jupiter and Saturn in low-eccentricity, low-inclination orbits slightly closer together than
their present locations, as predicted by the Nice Model \citep{nice1,tsiganis:2005,nice2} (hereafter termed `Circular Jupiter and Saturn', CJS)
and four with Jupiter and Saturn in their current, slightly eccentric orbits (hereafter termed `Eccentric Jupiter and Saturn', EJS). Multiple
terrestrial planets formed within 250Myr in all simulations. The general system structure is shown in Figure \ref{final_sys} and the planetary
orbital properties are shown in Table \ref{prop}.

It is currently unclear whether CJS or the EJS (or some other configuration) best represents the initial configuration of Jupiter and Saturn.
\citet{nice1}, \citet{tsiganis:2005}, and \citet{nice2} have shown how the CJS configuration of the giant planets, and their subsequent orbital
evolution via interactions with the primordial trans-Neptunian disk, can explain their current orbital configuration, the populations of Jupiter
Trojan asteroids, and the Late Heavy Bombardment of the inner Solar System.  \citet{morbidelli:2007} has shown that such a configuration is a
natural consequence of the giant planets' formation in, and interactions with, the protoplanetary disk.  However, simulations of terrestrial
planet formation tend to find that some aspects of the terrestrial planets, namely a compact system and a small Mars, are best reproduced with a
configuration more like in the EJS simulations, where Jupiter has an eccentricity comparable to or larger than its current value
\citep{dave,thommes:2008,raymond:2009}.  Given the current uncertainty, we study both the CJS and EJS simulations here, as they give a good
representation of the range of possible initial configurations for Jupiter and Saturn.

It should be noted that the formation of Mercury analogs is not currently possible in the simulations of \cite{dave} as all of the embryos begin
with a mass almost twice that of Mercury. Additionally, the planetesimals are not gravitationally interacting with each other, preventing
formation of a small `Mercury' via planetesimal accretion. Finally, perfect accretion is assumed throughout the simulations, preventing giant
impact events from striping off the crust and mantle of a differentiated proto-Mercury, a scenario that widely suggested to explain Mercury's
current high density. As the simulations, results and their implications have already been discussed in great detail by previous publications,
the reader is referred to \citet{dave} for further discussion of these aspects.

\subsection{Chemical \label{method}}
The chemical composition of material within the disk is assumed to be determined by equilibrium condensation within the primordial solar nebula.
This assumption is justified as numerous detailed analyses of primitive chondritic material have clearly shown that their bulk elemental
abundances are smooth functions of their equilibrium condensation temperature, as determined for a solar composition gas at low pressure
\citep{davis}. Several possible explanations for this pattern have been proposed, with the most widely accepted being that the chemistry of
their parent planetesimals was established by  midplane temperature and pressure profiles \citep{cassen01}. Alternatively, the sluggish nature
of reaction kinetics within the cooler regions of the outer nebula may also produce the observed trend. Although the exact cause of the
primitive chondritic pattern is still unclear, we are able to utilize the fact that equilibrium condensation temperatures are reliable proxies
for the bulk elemental abundances of rock-forming elements within the early Solar System. Furthermore, observational evidence of these
equilibrium compositions is still seen today, preserved as the thermal stratification of the asteroid belt \citep{gradie}. The mineralogy of
primitive chondritic meteorites is similar in most respects to that predicted by equilibrium condensation \citep{ebel}. Thus the assumption of
equilibrium composition is a valid starting point for this type of study. This assumption in turn implies that the primary controls on embryo
and planetesimal compositions are the radial midplane temperature and pressure gradients encountered by the nebular material.

In order to determine the equilibrium composition of the solid material, we utilized the commercial software package HSC Chemistry (v. 5.1). HSC
Chemistry determines the equilibrium chemical composition of a system by iteratively minimizing the system's Gibbs free energy, using the GIBBS
equilibrium solver, as described in \cite{white}. This software has been successfully used in recent studies of solar nebula chemistry
\citep{pasek} and supernova stellar outflows, producing compositions that correlate with observed mineralogy in presolar grains found in
interplanetary dust particles \citep{messenger}. As we are working with similar spatial, thermal and pressure scales as these previous studies
and as the key theory applied here is not spatially dependent, we feel confident in applying it to this current study. Several limitations have
been encountered with the program, primarily the limited number of elements and species possible to incorporate in a simulation and its
inability to handle species in non-ideal solid solutions such as pyroxene and olivine, two of the major rock-forming minerals on the terrestrial
planets. End member species can and have been considered but the non-ideal solid solution interaction between them is not currently possible.
These limitations will not significantly affect the final conclusions of this current study but will be the subject of future work. The list of
solid and gaseous species included in the HSC Chemistry calculations are shown online in Table \ref{HSC} (online supplemental material). Note
that although silicate liquids were found to be stable in dust-enriched systems \citep{ebelgross}, here we are only considering compositions
with a solar dust/gas ratio \footnote{Here the dust/gas ratio is taken to be the cosmochemical definition, that is the atomic Si/H ratio of
3.5$\times$10$^{-5}$. The usual astrophysical definition of the dust/gas ratio is the ratio of mass of the solid material to the mass of the gas
within the system. As expected, this value varies over the disk as condensation progresses.}. As such, liquids do not play an important role in
the condensation sequence and are thus neglected from the current calculations.

Our present simulations incorporate 14 major rock-forming elements (C, N, O, Na, Mg, Al, Si, P, S, Ca, Ti, Cr, Fe and Ni), along with H and He,
which dominate the partial pressures of protoplanetary disks. Current solar photospheric abundances are utilized as a proxy for the initial
composition of the solar nebula. All inputs are assumed to initially be in their gaseous, elemental forms. Solar elemental abundances for all
elements were taken from \cite{asp} and are shown in Table \ref{solar-inputs}. Although there has been some recent discussion about revising the
Solar C and O abundances \citep{socas}, we have adopted the most recent and widely accepted values for this current study. The 50\% condensation
temperature calculated for each element in a gas with a Solar photosphere composition is shown in Table \ref{cond}. The 50\% condensation
temperature is the temperature at which 50\% of the original element has condensed, with the remaining 50\% still in the gaseous phase. As such,
it provides an effective measure of the volatility of a given element in the system. It should be noted that although O is present in high
temperature condensates and silicate species, it does not obtain 50\% condensation until water ice condenses within the system, hence its low
50\% condensation temperature. Additionally, the 50\% condensation temperatures from the models of \cite{lodders} are also provided. The average
difference between this study and that of \cite{lodders} is just +11.75K (defined as T$_{\rm present\:study}$  - T$_{\rm Lodders}$), with the
largest difference being 83K for P. The excellent agreement between the temperatures obtained in this study and those of \cite{lodders} implies
that the chemical models utilized in this study are accurately reproducing the thermodynamical condensation sequence of the solar disk.

In order to relate the chemical abundances to a spatial location within the original disk, we used midplane pressure and temperature values
obtained from the axisymetric $\alpha$ viscosity disk model of \cite{hersant}. The \cite{hersant} model is a two-dimensional time-dependent
turbulent accretion disk model incorporating vertical disk structure, turbulent pressure and self-gravity. As for other standard disk models, a
Keplerian rotation law, hydrostatic equilibrium and energy balance between viscous heating and cooling due to radiative losses are assumed
\citep{hersant}. The effects of irradiation from the central star are neglected, as are other disk features such as inner holes and shadow
zones. In the current study, we limit the midplane conditions to be those produced by the ``nominal'' model defined by \cite{hersant} to be a
stellar mass accretion rate of 5$\times$10$^{-6}$M$_{\bigodot}$yr$^{-1}$, an initial disk radius of 17 AU and an $\alpha$ value of 0.009.
Previous work has successfully applied this model to constrain the bulk composition of Jupiter (e.g. \citealt{pasek}). Pressure and temperature
values were determined with an average radial separation of 0.03AU throughout the study region. Due to computational limitations, pressure
values were rounded to the nearest order of magnitude. This is not expected to greatly alter the final planetary compositions produced as
temperature exerts a far greater control on the equilibrium composition at a given radii.

The midplane temperature and the pressure, and thus also the equilibrium composition of solids present within the disk, changes with time as the
disk evolves. In order to capture this effect in our simulations, we determined an ensemble of predicted planetary compositions constrained by
disk conditions at multiple time periods. At the earliest, we use the temperature and pressure profiles obtained for disk conditions at t =
2.5$\times$10$^{5}$yr, where conditions were first determined to be suitable for solids to be present across the entire radial region being
modelled in the dynamic simulations. We end with disk conditions at t = 3$\times$10$^{6}$yr, the average lifetime for the protoplanetary gas
disks \citep{haisch}. Five cases between these end points are considered (for disk conditions at t = 5$\times$10$^{5}$yr, 1$\times$10$^{6}$yr,
1.5$\times$10$^{6}$yr, 2$\times$10$^{6}$yr, 2.5$\times$10$^{6}$yr). Thus we determine the planetesimal and embryo compositions for a total of
seven cases, covering the entire range of times during which embryos and planetesimals could potentially form, either early or late in the
lifetime of the disk.  Note that the timescale used for the evolution of the disk conditions is not coupled with the timescale of the dynamical
simulations. Rather, chemical compositions are simply determined for a range of disk mid-plane conditions and no time-variation in disk
conditions over the duration of the dynamical simulation is incorporated.

\subsection{Combining Dynamics and Chemistry\label{combo}}
In order to combine these two different modeling approaches, we assume that each planetary embryo and planetesimal of the dynamical models
retains the chemical composition in equilibrium with the nebula in the region that it first formed. The bulk compositions of the final planets
are simply the sum of each object they accrete. By tracing the origin of each embryo and planetesimal incorporated into the final planets of the
\citet{dave} dynamical simulations, and calculating the chemical composition of those bodies based on their original locations, we constrain the
bulk composition of the final terrestrial planets. This procedure was completed for all planets at each of the seven
different time steps simulated within the chemical models.

\subsection{Stellar Pollution}
Stellar pollution refers to the late addition of material to the stellar photosphere, usually during planet formation and/or migration. It has
been suggested as being a possible cause for the high metallicity trend observed in known extrasolar planetary host stars \citep{la,g6,murray}.
The extent of stellar pollution produced by terrestrial planet formation simulations of \cite{dave} was measured by determining the amount and
composition of material accreted by the Sun during the formation process and the resulting photospheric elemental abundance changes such an
addition would produce. Any solid material migrating to within 0.1AU from the Sun is assumed to have accreted onto the Solar photosphere through
gravitational attraction. This material is then assumed to have been uniformly mixed throughout the solar photosphere and convective zone.
Gravitational settling and turbulence within the convective zone are neglected from this present study, primarily due to the large time scales
involved in settling (e.g. $\sim$10\% reduction in photospheric heavy metal abundances by the current solar age \citep{pin}) and the
uncertainties involved in the effects of turbulence on specific elements. Such models are beyond the scope of this paper. However, it is
expected that each of these processes would reduce the amount of accreted material present within the upper layers of the Sun, thus reducing the
photospheric elemental abundance. As such, the resulting abundance changes determined in this study represent the maximum possible values that
we would expect to observe in the Solar spectrum. Additionally, at earlier times, the mass of the solar convective zone would have been greater
than the current value of 0.03 M$_{\bigodot}$, again resulting in a smaller change in photospheric abundance than predicted by this current
study.

The mass of each element accreted by the Sun was determined in the same way as described in section \ref{combo} for terrestrial planets. The
resulting photospheric abundance is given by:

\begin{equation}
{\rm [X/H]} = \rm log\left[\frac{f\rm_{X}}{f_{X, \bigodot}}\right]
\end{equation}\\

where [X/H] is the resulting abundance of element X after accretion of terrestrial planet material, f$_{\rm X}$ is the mass abundance of element
X in the Solar photosphere after accretion and f$_{\rm X, \bigodot}$ is the initial mass abundance of element X in the Sun before accretion
(from \citealt{murray}). f$_{\rm X, \bigodot}$ values were obtained by utilizing the solar abundances of \citet{asp} and a current solar
convective zone mass of 0.03M$_{\bigodot}$ \citep{murray}. The present approach only addresses pollution by the direct accretion of
planetesimals and embryos during the current simulations, yet pollution may also occur both before planetesimal accretion (and thus the
simulation) begins and through the accretion of dust produced by the impact and collisional events during the formation process. Pollution by
these processes is believed to be small, resulting in $\emph{at most}$ a factor of two increase in the amount of material added to the solar
photosphere.

\section{Results\label{sec:results}}
\subsection{Abundance Trends\label{trends}}
The bulk elemental abundances for all simulated terrestrial planets (as elemental wt\% of the planet) for each set of disk conditions examined are
provided online as supplemental material (Table \ref{SSwt}). Additionally, the Mg/Si value for each planet is also listed.

From these results, it can be seen that the bulk elemental abundances of the final terrestrial planets vary significantly with disk conditions,
especially for certain elements such as O, Al and Ca.  As expected, the relative abundances of the most refractory elements (Al, Ca, Ti) are
observed to decrease for disk conditions at later times while the more volatile elements (O, Na, H) all increase in abundance. The largest
variation occurs in the O abundance, with simulations for disk conditions at t = 3$\times$10$^{6}$yr producing terrestrial planets with up to
30.79 wt\% more O than identical simulations for disk conditions at t = 2.5$\times$10$^{5}$yr.

These compositional variations can be directly attributed to the changing conditions of the disk itself. As the temperature and pressure at the
mid-plane decrease with time, the equilibrium composition at a specific radius also changes. Thus the assumed composition of the planetesimals
and embryos also change, in turn producing variations in the final planetary abundances. Consequently, as time progresses and the disk cools,
the planetary composition can be seen to decrease in abundance of the most refractory elements (Al, Ca) and increase in the more volatile
elements (O, H) as volatile, hydrated species become stable throughout an increasingly  larger fraction of the disk.

\begin{figure}
\begin{center}
\includegraphics[width=120mm]{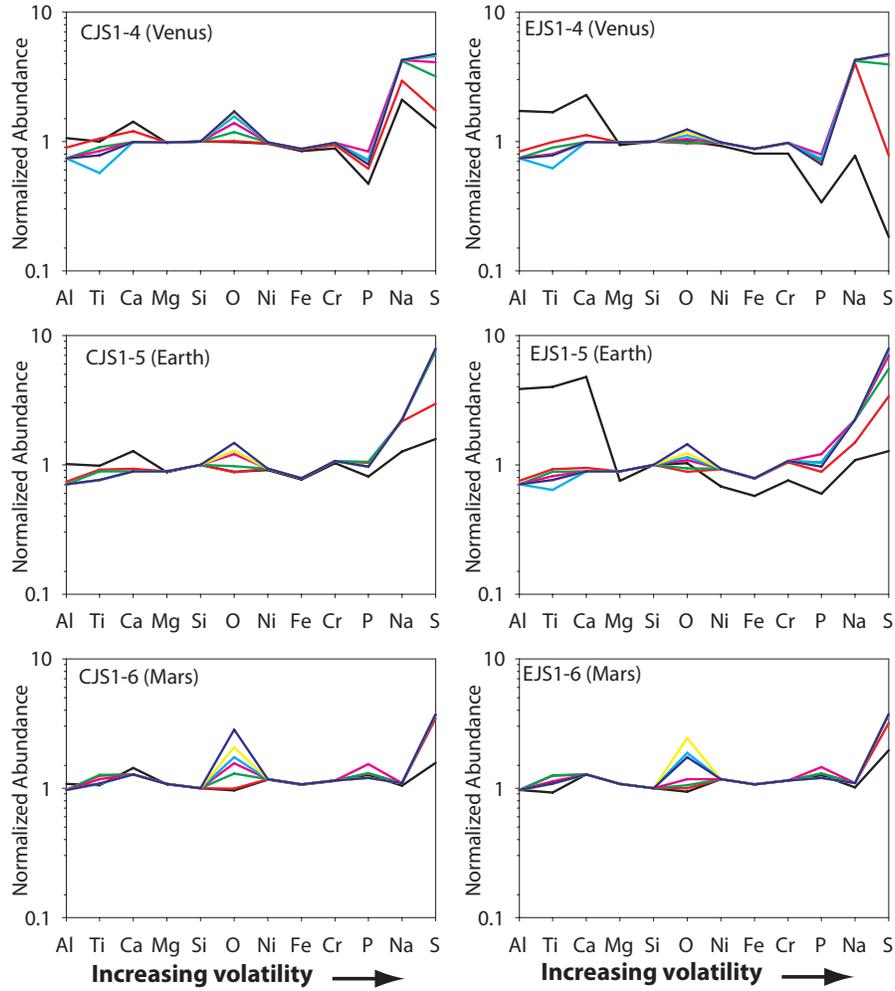} \caption[Normalized abundances for the CJS1 and EJS1 simulations.]{Normalized abundances for CJS1 and
EJS1 simulated terrestrial planets. The terrestrial planet each simulation is normalized to is shown in
parentheses. Values are shown for each of seven time steps considered with the following color scheme: black = 2.5$\times$10$^{5}$ years, red =
5$\times$10$^{5}$ years, green = 1$\times$10$^{6}$ years, pink = 1.5$\times$10$^{6}$ years, light blue = 2$\times$10$^{6}$ years, yellow =
2.5$\times$10$^{6}$ years and dark blue = 3$\times$10$^{6}$ years. \emph{Left:} CJS1 terrestrial planets. \emph{Right:} EJS1 terrestrial
planets. Reference Solar System planetary abundances were taken from \cite{manda} (Venus), \cite{kandl}(Earth) and \cite{landf}
(Mars).\label{spider1}}
\end{center}
\end{figure}

A more accurate picture of the predicted planetary abundances is obtained by examining the abundances produced for each specific set of disk
conditions simulated. Figure \ref{spider1} shows the Si and planet-normalized bulk elemental composition for the planets produced in the CJS-1
and EJS-1 simulations for all seven disk conditions. The simulated planets were normalized to the bulk elemental composition of Venus, Earth or
Mars on the basis of their orbital properties, primarily their semimajor axis. Reference Solar System planetary abundances were taken from
\cite{manda} (Venus), \cite{kandl}(Earth) and \cite{landf} (Mars). The normalized values were obtained via:

\begin{equation}
{\rm Normalized\:abundance\:of\:element\:X} = \rm
\frac{\left(\frac{wt.\:\%\:of\:X}{wt.\:\%\:of\:Si}\right)_{simulated\:planet}}{\left(\frac{wt.\:\%\:of\:X}{wt.\:\%\:of\:Si}\right)_{terrestrial\:planet}}
\end{equation}\\

Values above unity indicate an enrichment in the simulated terrestrial planet, while those below it indicate depletion.

All three suites of planetary reference values are themselves based (to some degree) on simulations. Bulk Earth values were obtained by
extrapolation of Best Bulk Silicate Earth (Best BSE or BBSE) abundance values (as obtained from previously published studies of mantle
xenoliths, pyrolite and basaltic elemental ratios), the observed volatility trend for volatile lithophile elements and the known physical and
seismic properties of the interior of the Earth \citep{kandl}. This approach, although necessary, does introduce uncertainties especially when
calculating the composition of the core as partioning ratios are not well known for all elements. Furthermore, the use of a volatility trend
requires knowledge of the condensation temperature of the elements. While such information is well known for several species (such as Fe, Mg and
Si with errors on the order of 3\% of the condensation temperature), larger degrees of uncertainty remain for more volatile elements. This
uncertainty produces errors in the predicted condensation temperatures of $\pm$14\% of the condensation temperature itself and results in final
elemental abundance errors of 5-10\% of the actual abundance for most elements, reducing to less than 1\% for Fe, Mg and Si. Of the elements
considered in the current study, those with the greatest overall uncertainty are O, S and C. Although not quantified by \cite{kandl}, the
uncertainty in O is derived from the fact that the O abundance was obtained based on the valence states of cations. The approach taken by
\cite{kandl} in obtaining their bulk compositional values is predicated upon the assumption that 15\% of all Fe present is in the form
Fe$^{3+}$. As such, any variation within the assumed valence state distribution will induce a variation in the O abundance of Earth. C and S
abundances similarly suffer from large uncertainties ($\sim$25\% of the given abundance) due to variations in the published BSE values, the
predicted condensation temperatures and the core partitioning coefficients.

Bulk Mars abundances were taken from \cite{landf} and are based on the combination of known meteorite material (H, CV and CI meteoritic
material) required to reproduce the oxygen isotopic abundances observed for the SNC meteorites. Variations in the elemental compositions within
each class of meteorite are thus the main source of uncertainty in the current estimates. \cite{landf} estimate their final abundance errors to
be $\pm$10\% for all elements studied.

Finally, the marked lack of data regarding the elemental abundances of Venus resulted in the necessary use of the purely theoretical bulk
planetary abundances of \cite{manda} for this study. \cite{manda} adopt a similar approach to the one used here and assume that the composition
of solid material initially present within the system is controlled by equilibrium condensation. The exact predicted composition was obtained
from the Ganapathy-Anders 7-component model \citep{morgan}. This model determines the composition of a body based on the amount of early
condensate (i.e. first solids present within the disk), metal, silicate, troilite, FeO, MnO and Cr$_{2}$O$_{3}$ believed to be incorporated into
a final planetary body. Bulk elemental abundances are thus inferred from general cosmic proportions as compared to an ``index element'' for each
component. As such, general and solar elemental ratios are assumed to be homogenous and remain constant for the entire planetary system. For the
simulations of \cite{manda}, the limited information available on the index elements for Venus was supplemented by geochemical restrictions from
the Earth, Moon and chondritic meteorites to obtain predicted bulk planet abundances. As such, large uncertainties exist in all elements
studied. As quantified uncertainty factors are not provided by \cite{manda}, it is impossible to gauge the errors on their predicted abundances.
However, they are believed to be considerable, resulting in elemental abundances for Venus that should be taken as a guide only.

From Figure \ref{spider1} it can be seen that although no single simulation exactly reproduced the compositions of the terrestrial planets, the
time when the adopted disk conditions produced simulated compositions that most closely resemble those of the terrestrial planets (the `best
fit' time) is 5$\times$10$^{5}$yr, based on the comparatively small deviations produced in the bulk elemental abundances. For disk conditions at
later times, significantly larger enrichments can be seen in O, Na and S for all planets. Although not shown here, disk conditions at this same
time also produces the best agreement between simulated and observed abundances for the other six simulations studied and as such future
discussions will focus on the compositions produced by the disk conditions at this time. This result does not imply that the material which
formed the terrestrial planets actually condensed out of the Solar nebula exclusively at 5$\times$10$^{5}$yr. Rather it is simply the time at
which the disk conditions and resultant snapshot of the chemistry of the Solar disk (as used here) best reproduced the expected abundances.
Similar plots for the other six simulations studied are included as supplemental material online.

The elemental wt\% produced for each of the simulated planets based upon disk conditions at 5$\times$10$^{5}$yr are shown in Table \ref{5E5}.
Numerically, these abundances agree with Solar System values to within 1 wt\% for all elements studied except for Mg (up to 2 wt\% deviation),
Fe and O (up to 2.5 wt\% deviation) and S (up to 5 wt\% deviation). In terms of normalized values, the predicted abundances are within a factor
of 0.5 for almost all elements studied. Enrichments can be seen in Na (up to 4$\times$ the expected value) and S (up to 8$\times$ the expected
value) and are discussed below. These abundances match well with observed Solar System values, indicating that
these terrestrial planet formation models are producing planets with both orbital properties and bulk elemental abundances
comparable to those of the terrestrial planets of the Solar System.

Despite the excellent agreement between the observed and predicted abundances, the planet-normalized spider plots of Figure \ref{spider1}
clearly display several systematic deviations from the expected terrestrial planet abundances. Specifically, considerable deviations can be seen
in P (Venus analogs only), Na and S for all simulations. This is partly due to the uncertainties in the expected planetary abundances for these
elements as was previously discussed. Additionally, as P, Na and S are some of the most volatile elements in this system, it is believed that
the observed enrichments are a result of the fact that we do not consider volatile loss during the accretion process. Both the dynamical and
chemical simulations assume that in any given impact, perfect merging of the two bodies occurs and all mass is retained. However, such violent
impacts are known to both melt and eject a considerable portion of both the target body and the impactor (such as in the moon forming impact
\citep{moon1,moon2}). This has the potential to drastically alter the bulk elemental abundances of a planet. It is expected that a significant
fraction of the most volatile species in the system will be lost from the planet to the surrounding disk during the accretionary process, thus
reducing the observed enrichments. This effect has not been incorporated into our simulations, resulting in volatile enriched final planetary
bodies. The results of a first order approach to include the abundance effects of volatile loss are discussed in Section \ref{loss}.

Some degree of radial compositional variation is captured by the current models. Simulated planets deemed to be Venus and Earth analogs were
found to produce equivalent normalized abundances for the major rock forming elements when normalized to the elemental abundances of Venus and
Earth. Differences in the quality of the fit occur for P, Na and S and are due to uncertainties in the abundances of Venus. The Mars analogs are
found to have a $\emph{slightly}$ better fit when normalized to the elemental abundances of Mars, as compared to those of Venus and Earth. This
difference is minor with Earth abundances arguably producing an equivalent normalization. However, given that the radial compositional
differences observed for Venus, Earth and Mars are small and that planetary accretion is stochastic and includes a significant amount of radial
mixing of material, it is not surprising that the approach adopted in this study has been unable to complectly reproduce the expected radial
variations.

\subsection{Planetary Geochemical Ratios}
\begin{figure}
\begin{center}
\includegraphics[width=100mm]{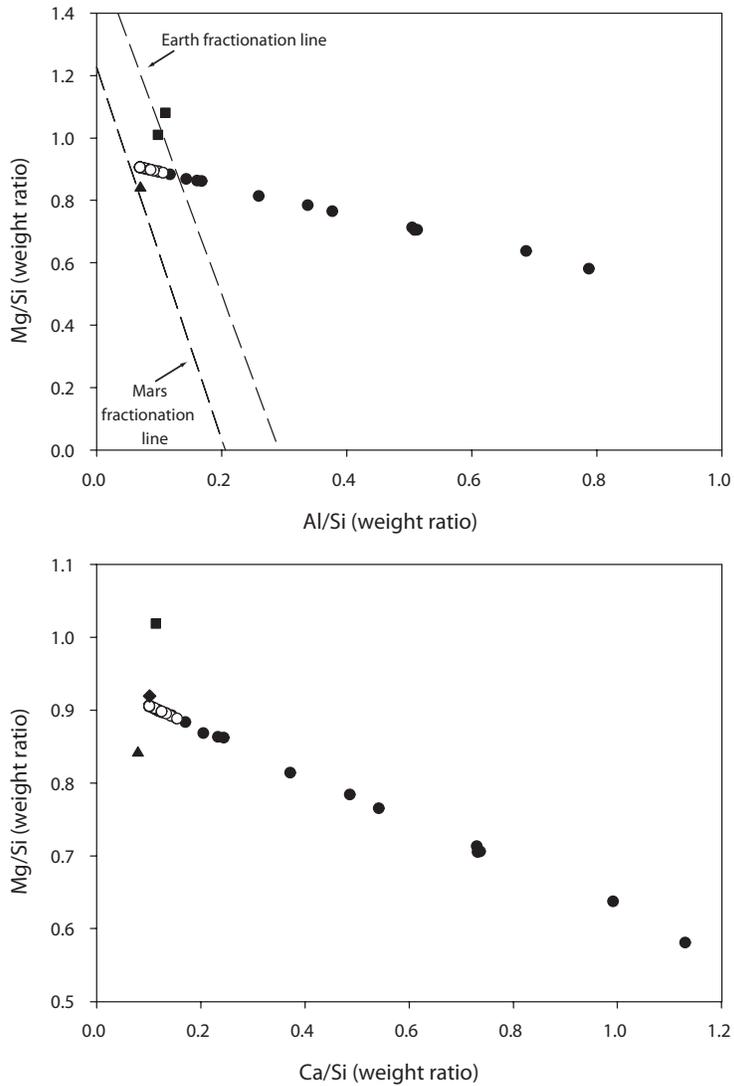}\caption[Cosmochem ratios for all Solar System simulated planets.]
{Cosmochemical ratios for all simulated terrestrial planets. \textit{Top}: Al/Si v. Mg/Si for all simulated terrestrial planets. \textit{Bottom}: Ca/Si v. Mg/Si for all simulated terrestrial planets. Filled circles indicate values for disk conditions at t = 2.5$\times$10$^{5}$ years while
open circles indicate values for disk conditions at t = 5$\times$10$^{5}$ years. Values at all other times are concentrated at the
5$\times$10$^{5}$ years values and are not shown for clarity. Earth values are shown by filled squares and are taken from \cite{kandl} and \cite{mands}. Martian values are shown by a filled triangle and are taken from \cite{landf}. Venus values are shown by a filled diamond and are taken from
\cite{manda}.\label{ratios}}
\newpage
\end{center}
\end{figure}

In addition to matching the bulk elemental abundances, simulations such as these should also reproduce the key geochemical ratios for the
terrestrial planets. Here we have considered the values of Mg/Si, Al/Si and Ca/Si. All three of these values differ by chemically significant
amounts from the expected planetary values for all planets produced. For simulations run with disk conditions at times of 5$\times$10$^{5}$
years and beyond, the planetary ratios are identical to the solar input ratios. This effect is produced by the fact that for disk conditions at
t = 5$\times$10$^{5}$ years, all the Mg, Al, Ca and Si has condensed out over the study region. This results in the solid species (and thus also
the planets produced) possessing the same chemical ratios as present in the initial solar nebula. For example, for \cite{hersant} disk
conditions at t = 5$\times$10$^{5}$ years, Mg/Si, Al/Si and Ca/Si ratios in the solid material reach Solar values by 0.5AU, clearly interior to
the primary feeding zone for the formation of Earth and Mars. Unfortunately, both the Mg/Si and Al/Si values are lower than is observed for the
Earth but above current Martian values while the Ca/Si values are above those of both Earth and Mars (but in agreement
for Venus) (see Figure \ref{ratios}). Thus for the majority of disk conditions presently considered we are producing planets that have
compositional ratios in between those of Earth and Mars and are slightly enriched in Ca.

Variations in these elemental ratios from solar ratios are produced in simulations based on disk conditions at 2.5$\times$10$^{5}$ years (see
Figure \ref{ratios}). At this time, the inner disk (within approximately 1AU) is dominated by Al and Ca rich species such as
spinel (MgAl$_{2}$O$_{4}$) and gehlenite (Ca$_{2}$Al$_{2}$SiO$_{7}$), resulting in planets with high Al/Si and Ca/Si values and low Mg/Si values
as can be seen in Figure \ref{ratios}. Thus it can be seen that there is a temporal variation in the chemical ratios of the
planets produced. This further supports our choice of disk conditions at t = 5$\times$10$^{5}$ years as producing the ``best fit'' abundances as
although the ratios discussed here do not precisely agree with the observed values, the deviation is significantly smaller than for earlier disk
conditions.

The fact that the simulated chemical ratios do not agree with observed planetary values suggests two possible causes. The first is that the
dynamical simulations are not forming planets from material sourced from the correct region of the disk. There is a small radial region within
the disk where the values of Mg/Si, Al/Si and Ca/Si are in agreement with those of Earth. This region occurs between 0.61 to 0.68AU
(corresponding to temperatures of 1352 and 1305K) for disk conditions at t = 2.5$\times$10$^{5}$ years and 0.12 to 0.13AU for disk conditions at
t = 3$\times$10$^{6}$ years for the disk model of \cite{hersant}. Thus one possible solution for the current difference may be that the
dynamical models need to form Earth from material located within this region. However, it is exceedingly difficult to imagine how such a
formation process would occur as it requires the movement of a large amount of material over a significant distance through the disk. Current
formation simulations do not produce this degree of radial mixing and furthermore it is questionable whether sufficient material to form the
terrestrial planets would even be located within this region, making it unlikely that such a scenario is feasible. Additionally, bulk elemental
planetary compositions produced at these temperatures would not be in agreement with observed bulk planetary elemental abundances. Finally,
observations have suggested that the temperature at 1AU in the disk of young stellar objects is less than 400K \citep{beckwith}, far less than
the $~$1300K required to produce the necessary Mg/Si, Al/Si and Ca/Si values at 1AU. This suggests that the dynamical models utilized here are
not the source of the deviation in Mg/Si, Al/Si and Ca/Si values.

The second possibility is that the composition of solids within the disk are controlled by disequilibrium condensation. Disequilibrium
condensation refers to the chemical model in which once a solid has condensed, it is removed from the system and may no longer interact with the
remaining gas. Such a process may occur if condensation occurs rapidly enough to for large bodies to grow quickly and thus shield the interiors
from further equilibrium reactions \citep{cowley}. Under such conditions, secondary condensates would have Mg/Si values above average
\citep{sears}, thus possibly increasing the Mg/Si value to the required Earth value over the primary feeding zone in the \cite{dave}
simulations. The possible role of disequilibrium processes does not invalidate our initial assumption of equilibrium controlled abundances. The
assumption of equilibrium is acceptable for determining bulk elemental abundances but finer details of planetary composition will clearly need
to incorporate a greater interaction between a variety of equilibriums and disequilibrium processes. Furthermore, equilibrium-based predictions
for simulations of this type need to be calculated initially to provide a baseline for future disequilibrium studies.

\subsection{Oxidation State}
\begin{figure}
\begin{center}
\includegraphics{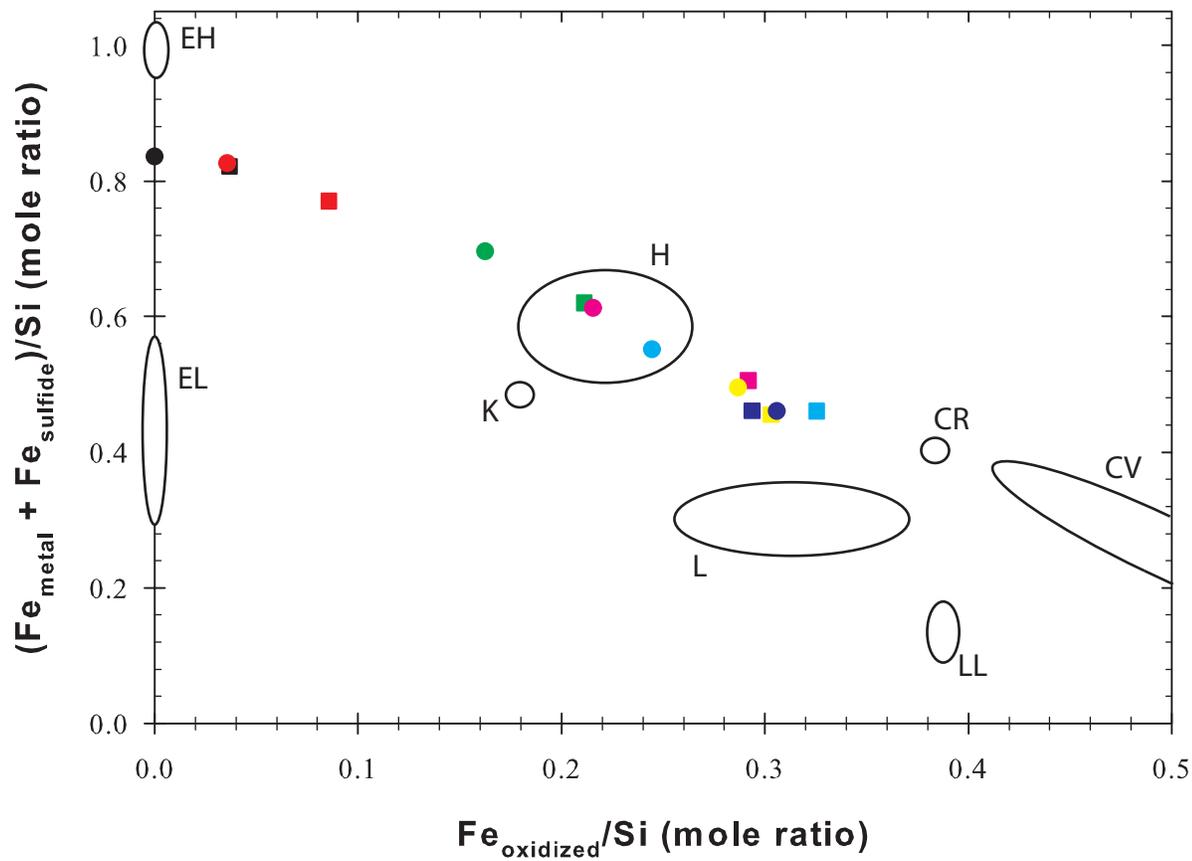}\caption[Oxidation state plot for CJS1 and EJS1 simulated planetary abundances.]
{Urey-Craig oxidation state plot for CJS1 and EJS1 simulated planetary abundances. Squares indicate CJS1 values. Circles indicate EJS1 values. Values are
shown for each of seven time steps considered with the following color scheme: black = 2.5$\times$10$^{5}$ years, red = 5$\times$10$^{5}$ years,
green = 1$\times$10$^{6}$ years, pink = 1.5$\times$10$^{6}$ years, light blue = 2$\times$10$^{6}$ years, yellow = 2.5$\times$10$^{6}$ years and
dark blue = 3$\times$10$^{6}$ years. Approximate regions for several meteorite groups are also shown.\label{redox}}
\newpage
\end{center}
\end{figure}

The oxidation state of the planets is of great chemical and geological interest. As we are currently calculating predicted bulk elemental
abundances and not detailed mineralogies for the simulated planets, estimates of the bulk oxidation state of the planet are obtained by
calculating the oxidation state of the embryos and planetesimals before accretion occurred based on their equilibrium compositions. The
resultant oxidation states are shown in Figure \ref{redox}. The current simulations are producing very reduced planetary compositions for the
earliest set of disk conditions studied, containing large amounts of metallic Fe and little FeO. For disk conditions at later times, the
compositions become more oxidized, evolving through the oxidation state of the H-type meteorites to finish closer to the redox state of the CR
meteorites, primarily due to the increased amount of magnetite (Fe$_{3}$O$_{4}$) and fayalite (Fe$_{2}$SiO$_{4}$) accreted by the planets. For
midplane conditions at t = 2.5$\times$10$^{5}$ years, fayalite is only present in significant amounts beyond 1.07AU while magnetite is only
present beyond 4.82AU, beyond the dynamical simulation range. For conditions at t = 3$\times$10$^{6}$ years, fayalite is present beyond 0.2AU
with magnetite now present beyond 0.95AU. Thus for later disk conditions, the feeding zones of the terrestrial planets (and thus the planets
themselves) are more oxidized.

In addition to the observed temporal variations seen in the current model, there is some evidence to support the idea that the localized
oxidation state varied throughout the solar disk. For example, \cite{devin} outlines how the type I chondrules present in CR chondrites formed
in a reducing environment while the type II chondrules (also present in CR chondrites) formed later under oxidizing conditions. However, these
variations in the ambient oxidation state over potentially small scales (such as the formation region of CR chondrites) are not incorporated
into the current approach. Similarly, the dust/gas ratio is also well known to alter the oxidation state of a system and the stable condensates
\citep[see][and references therin]{ebelgross}. Such variations in the oxidation state may prove to be necessary to produce the parent bodies of
certain meteorite classes, such as the enstatite meteorites, which are not produced under the current equilibrium assumption. Enstatites contain
a very reduced mineralogy, suggesting a C/O value above solar \citep{larimer:1979}. Such regions and localized variations are neglected from the
current approach as they are not yet well understood or simulated. The migration of material within the disk can also act to alter the oxidation
state of the solid material. For example, the redistribution of water within the inner 5AU of the disk over time via diffusion and advection has
been found to significantly alter the oxidation state of the disk itself \citep{pasek,cyr}, producing both reducing and oxidizing regions of
various widths and locations (depending on the initial conditions). As a result, the redox state of the solid material should be drastically
altered as the disk evolves. Until we are better able to model these variations, we are unable to examine their effects on the final simulated
planetary compositions produced. It is also worth mentioning that the extremely reduced nature of the solid material may also be increased to
some extent by our present inability to simulate the olivine and pyroxene solid solutions. This acts to limit the amount of Fe that can be
oxidized and incorporated into silicate species, thus producing more reduced compositions.

Comparison of simulated oxidation values to those of the terrestrial planets is difficult, either due to lack of direct information (Venus) or a
varying redox state (Earth and Mars). Earth has a reduced metallic core and more oxidized mantle (near QFM) \citep{jones,rollinson}. Although
the SNC meteorites span a large range of oxidation states (IW to IW+2.5\footnote{IW refers to the iron-wustite buffer. IW+2.5 refers to an
oxygen fugacity 2.5 log units above that of the buffer.}), the martian interior itself may have an oxidation state close to IW$-$1
\citep{jones}. Defining the "average" oxidation state of the Earth is complicated by variations due to tectonic settings, especially for the
crust and uppermost mantle \citep{rollinson}. Furthermore, it is believed that the redox state has varied through time (e.g. \citealt{galimov}).
In order for core formation to occur, a mantle composition of IW-2 to IW-3 is required \citep{jones}. However, the mantle must have achieved its
present oxidation level ($\sim$QFM) relatively early in the Earths history as little temporal variation is observed in the oxidation state of
basalts younger than $\sim$3.5 - 3.9 billion years \citep{delano}. As a result of these issues, there is currently no way of accurately
determining an oxidation state for the primitive terrestrial planets (and thus the time of the `best fit' disk conditions implied by it) at the
stage of formation that we are simulating here. However, the generally reduced nature of the planets produced in the current simulations is in
agreement with the hypothesis of an initially reduced Earth and Mars before planetary processes altered the redox state of their upper layers.

\subsection{Orbital properties of Jupiter and Saturn}
Although differences in the orbital and dynamical properties of the final planets are produced by the two different types of simulations
considered here \citep{dave}, the differences between the CJS and EJS simulations in terms of both their bulk elemental abundances and their
geochemical ratios for rock forming elements are negligible. Neglecting the small volatile rich planet formed in the asteroid belt in simulation
EJS3, comparable bulk compositional trends can be seen in both the CJS and EJS simulations. This clearly indicates that in the case of
terrestrial planet formation within the Solar System, the bulk composition of the final terrestrial planets formed is not highly dependent on
the orbital properties of Jupiter and Saturn. Difference between the two types of simulation emerge when we consider the delivery of hydrous
species (Section \ref{hyd}).

Once again, this is a result of the fact that the majority of planet forming elements have fully
condensed out of the nebula by $~$1200K, corresponding to midplane radii within 0.75AU from the host star for the models of \cite{hersant} and do
not change greatly in their relative weight percentage values over the remainder of the simulation region. As a result the majority of the solid
mass within the present simulations has solar elemental abundances. This is consistent with the idea that the equilibrium chemical composition
of the Solar nebula is not highly zoned in that it does not contain radially narrow zones of vastly different compositions. Instead, the vast
majority of the solar disk is dominated by material composed of pyroxene (MgSiO$_{3}$), olivine (Mg$_{2}$SiO$_{4}$) and metals (Fe and Ni).

\begin{figure}
\begin{center}
\includegraphics[width=120mm]{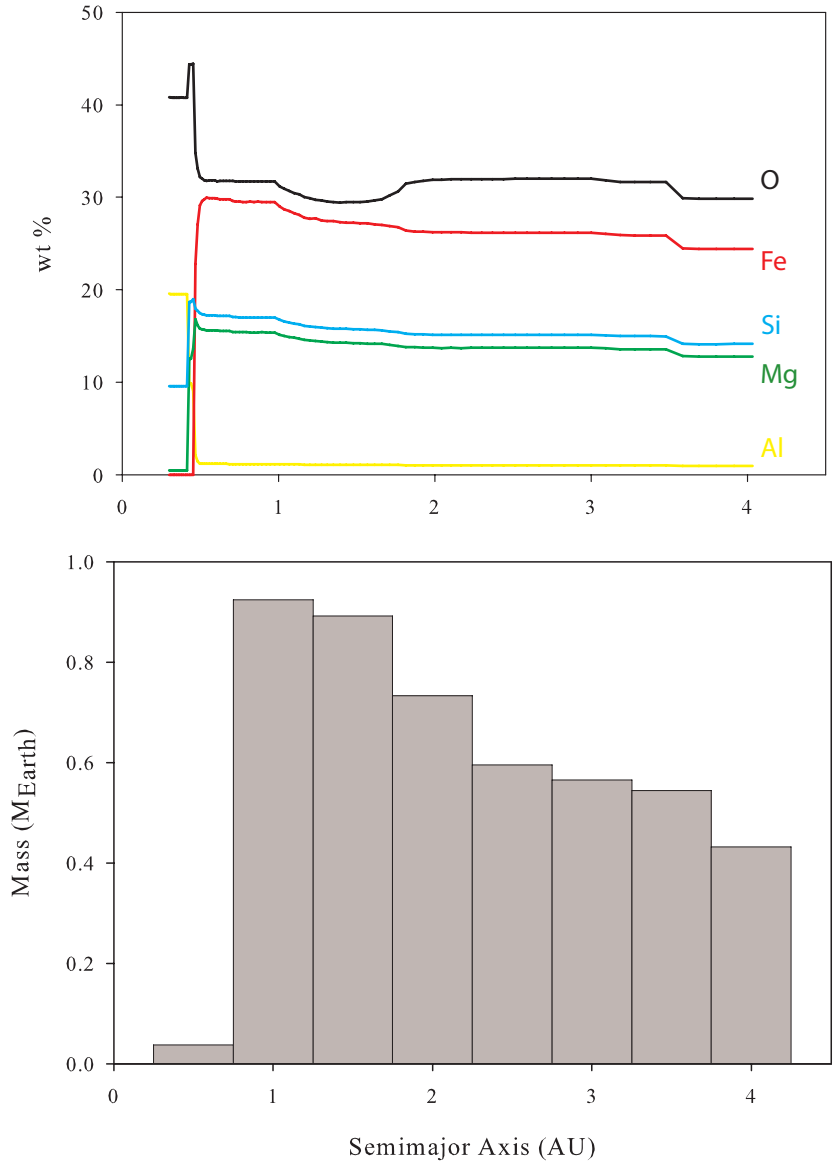}\caption[Distribution of mass and its composition.]{Distribution of solid mass and its relative composition within the Solar
disk at 5$\times$10$^{5}$ years. $\emph{Top:}$ Composition (in wt\%) for the solid material within the disk for O, Fe, Mg, Si and Al. $\emph{Bottom:}$ Initial distribution of mass within the dynamical simulations of \cite{dave}.\label{dist}}
\end{center}
\end{figure}

Figure \ref{dist} shows the weight percentage (in solid material) of the key planet building elements O, Fe, Mg and Si, in addition to the mass
distribution of the dynamical simulations. It can be seen that the majority of mass is expected to have similar relative elemental abundances.
The small variations observed in the weight percent values are caused by increases in the solid portion of disk mass generated by the
condensation of FeS at 1AU and serpentine (Mg$_{3}$Si$_{2}$O$_{5}$(OH)$_{4}$) at 3.5AU. Thus small variations in the feeding zones produced by
different orbital properties of Jupiter and Saturn within the CJS and EJS simulations are unlikely to produce major differences in final bulk
composition.

\subsection{Variations with Time \label{timing}}
\begin{figure}
\begin{center}
\includegraphics[width=120mm]{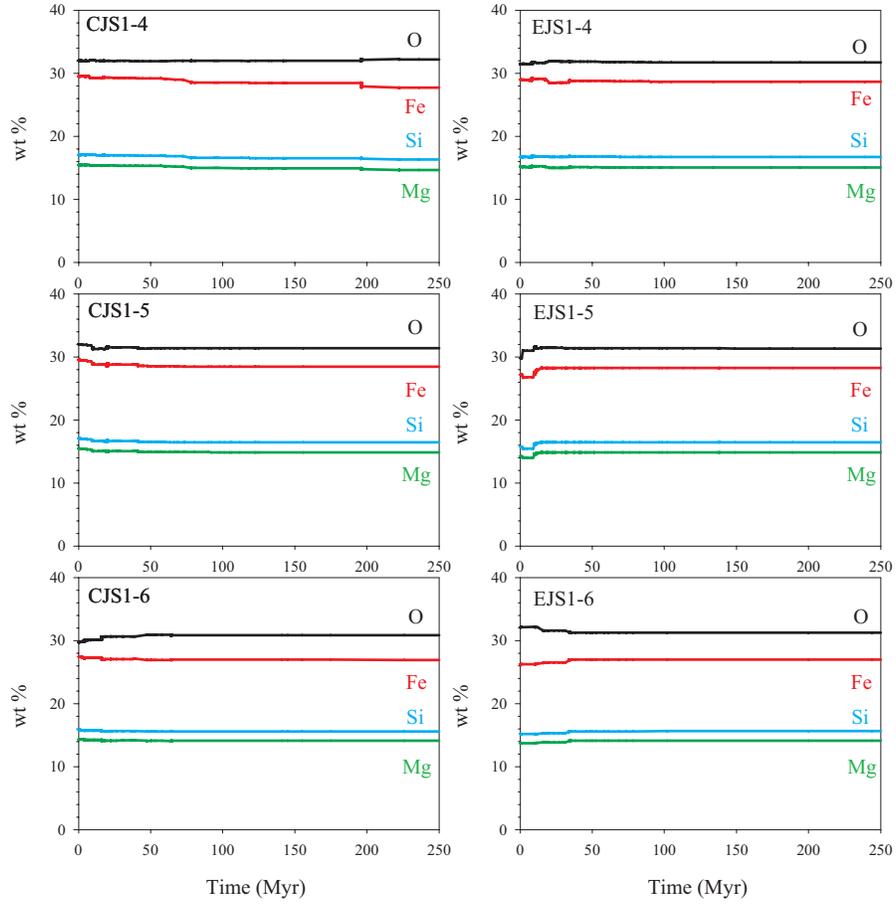} \caption[Variation in composition with time for the final planets produced by the CJS1 and EJS1 simulations.]
{Temporal variation in the elemental abundances of the final terrestrial planets produced by the CJS1 and EJS1 simulations. Variations in
compositions are due to the accretion of embryos and planetesimals throughout the dynamical simulations. \emph{Left:} CJS1 simulation results.
\emph{EJS1:} EJS1 simulation results.\label{time}}
\end{center}
\end{figure}

Figure \ref{time} shows the variation with time for weight percentage values of several key planet building elements in the planets produced by
the CJS1 and EJS1 simulations. It can be seen from Figure \ref{time} that very little variation in composition ($<$5 wt\%) occurs during the
formation process with the bodies obtaining their final compositions relatively early during accretion and displaying only minor deviations over
time. The same trend is also observed for the six other simulations not shown. This implies that the planets within these simulations formed
homogenously (i.e. from material with similar composition to the final planet). If formation did occur in this manner, then the presence of a
800-1000km deep magma ocean would be required during core formation in order to produce the siderophile element abundances observed in the crust and
upper mantle of the Earth \citep{drake}.

However, it is likely that the homogenous accretion observed here is due to the `snapshot' approach we have taken when determining the
composition of solid material within the disk. In these simulations, we have considered the composition determined by disk conditions at just
seven discrete times. In reality, it is quite likely that the composition of solid material will change over time as it undergoes migration and
experiences other disk and stellar processes (such as the redistribution of water \citep{cyr}). Such changes are not captured in our current
approach. As such, more detailed chemical models incorporating temporal variations in the composition of solid material are required in order to
test the homogenous accretion produced here.

\subsection{Late Veneer}
An alternative hypothesis to homogenous accretion is heterogeneous accretion where the composition of material accreted changes significantly
with time. Under this hypothesis, the highly siderophile element distribution within the Earth is explained via accretion of a ``late veneer''
\citep{chou}. Referring to the last $\sim$1\% of mass accreted by the Earth, this material would need to contain chondritic abundances of the
siderophile elements, and essentially no metallic iron \citep{dandr}. Although the CI and CM carbonaceous chondrites are sufficiently oxidized
and contain limited metallic Fe, their Os isotopic abundances are not in agreement with values required for the late veneer. On the other hand,
H-chondrites do posses Os isotopic ratios of the correct value but are not sufficiently oxidized. As such, we currently have no samples in the
meteorite record that represent a possible source for the late veneer material (\citealt{dandr}). However, this is not surprising, given the
limited number of chondritic parent bodies in our meteorite collection ($\sim$15) compared to the thousands of planetesimals that accreted to
form the Earth.

As expected from the previously observed homogenous accretion, the late veneer of the present simulations is similar in composition to the final
planetary abundance. Here the late veneer is taken to be the material accreted after the last impact by a projectile with mass $>$ M$_{embryo}$.
This material is not highly oxidized and is primarily composed of olivine, pyroxene, metallic iron, troilite (FeS), diopside
(CaMgSi$_{2}$O$_{6}$), nickel and albite (NaAlSi$_{3}$O$_{8}$). This composition is similar to that of the ordinary chondrites. Furthermore, as
noted in \cite{dave}, the planets produced in the EJS simulations accrete an average of $\sim$10\% of the final planetary mass as a late veneer,
an order of magnitude above the predicted amount. Thus it can be seen that the current simulations do not successfully reproduce the late veneer
material. However, as discussed in sections \ref{trends} and \ref{timing}, migration of material and temporal variations in the composition of
solid material accreted have the potential to drastically alter the composition of the late veneer. Such variations are not currently captured
by the current simulations.

\subsection{Hydrous Species\label{hyd}}
Differences between the CJS and EJS simulations emerge when considering the delivery of hydrous species to the final planets. In the present
simulations, ``hydrous species'' refers to water ice and the aqueous alteration product serpentine (specifically clinochrysotile
Mg$_{3}$Si$_{2}$O$_{5}$(OH)$_{4}$). Although silicate melts on the bodies themselves may contain water (up to 15 wt\% \citep{hamilton})and thus
also be an important source of hydrous material, they are currently neglected from this study as we are only focussing on the equilibrium
condensation composition of the system. As stated in \citet{dave}, the planets formed in the EJS simulations do not accrete significant amounts
of volatile rich material from beyond 2.5AU (with the exception of the same planet formed in the EJS3 simulations as previously mentioned). Thus
for disk conditions before 1$\times$10$^{6}$ years, these planets are found to contain no hydrous species. For disk conditions after this time,
however, all terrestrial planets produced in the EJS simulations are found to have significant water components, although generally less than
those of the CJS terrestrial planets. The appearance of water and serpentine at later times is largely due to the migration of the ice line to
within the region from which source material is obtained. The planets formed via the CJS simulations, however, all contain significant amounts
of serpentine and water ice for disk conditions at t = 5$\times$10$^{5}$ years. At this time, seven planets include hydrous material
($<$0.008M$_{\bigoplus}$), incorporated via the accretion of one to four planetesimals each containing a minor amount of serpentine. Only one
planet contained a large amount of serpentine ($>$0.01M$_{\bigoplus}$). All eight planets accreted the hydrous material before the late veneer
and during the stage where large, violent impacts were occurring. As such, it is expected that a significant component of this hydrous material
would be vaporized during later impact events and subsequently lost from the final planetary body or incorporated into the planetary core. Only
four planets (including two Earth analogs) accreted serpentine (0.001 - 0.003M$_{\bigoplus}$) as part of their late veneer. Given the relatively
late delivery of this material, it is believed that the majority would be retained during later planetary processing. As one would intuitively
expect, the amount of hydrous material accreted by a planet increases with increasing planetary semimajor axis.

Assuming that all of the hydrogen accreted as hydrous species is converted to water, the current CJS simulations are producing planets
containing 0.6 to 24.8 Earth ocean masses of water for disk conditions at 5$\times$10$^{5}$ years\footnote{1 Earth ocean mass of water =
1.4$\times$10$^{21}$kg = 2.34$\times$10$^{-4}$M$_{\bigoplus}$}. If we assume that Venus initially possessed a similar amount of water to the
Earth while Mars contained 0.06 - 0.27 Earth oceans of water \citep{lunine}, then it can be seen that for all of the terrestrial planets
considered, the present simulations are producing planets with sufficient water to avoid the need to invoke other large-scale exogenous water
delivery sources such as cometary impacts. Of course, these values should be considered to be upper limits on the amount of total water present
within the planet as potentially large amounts ($>$50\%) may be lost during the accretion process \citep{canup}. Additionally, some will
undoubtedly be lost  by photodissociation, Jean's escape and (definitely in the case of Earth) by the formation of organic species. Primordial
accretion of water (as predicted here) would result in H being accreted to the core of the planet via increased H partioning into molten FeHx
from a hydrated silicate melt \citep{okuchi2,okuchi}. This would again act to decrease the amount of water available on the planetary surface.
Although values for the amount of water expected to be lost by these processes are unknown, it is still likely that a significant portion of the
initial water will be retained within or on the crust. The presence of water would also have a variety of effects on the nature of the
proto-earth. It would alter the phase relations for the key components of a magma ocean, modify the physical properties of the melt (such as its
compressibility) and also the crystal settling properties of such a melt \citep{abe}.

Temporal variations in the composition of the disk are likely to increase the amount of hydrous material accreted by the planets due to
water-rich material being stable over a drastically larger fraction of the disk. These time-dependent variations can be observed in our current
approach as simulations undertaken for conditions at 3$\times$10$^{6}$ years produce terrestrial planets containing up to 1200 Earth ocean
masses of water, well above the levels observed for simulations under disk conditions at 5$\times$10$^{5}$ years. The current approach also does
not account for the possibility of water delivery via adsorption of water onto solid grains later incorporated into planetesimals
\citep{dandc,stimpfl}. As all of the solid material considered in the current simulations would be bathed in H$_{2}$O vapor, it is possible that
a significant amount of water could be delivered to the final planets via adsorption that is not presently accounted for. Thus it appears likely
that the terrestrial planets form ``wet'' with a sizeable portion of their primordial water delivered as a natural result the planetary
formation process.

This conclusion does not consider the resulting D/H ratio of the accreted water, a key constraint on any water delivery hypothesis, as isotopic
abundances can not be determined in our current approach. Primordial D/H ratios are known to vary greatly, from 2$-$3$\times$10$^{-5}$ for
protostellar hydrogen \citep{lecluse} to 9$\times$10$^{-5}$ for aqueous inclusions in meteorites \citep{deloule}. Similar variations can be seen
in solid bodies with carbonaceous chondrites containing values close to that of the Earth (1.5$\times$10$^{-4}$) while Mars appears to be
enriched in D with D/H values around 3$\times$10$^{-4}$ \citep{dandr}. Thus it is difficult to make detailed predictions about the possible D/H
ratios of the simulated planets. However zeroth order predictions can be made if we assume that the D/H ratio decreases linearly with increasing
semimajor axis from the observed Martian value to that of Jupiter's atmosphere (2.1$\times$10$^{-5}$)\citep{robert}. As all of the hydrous
species in the current simulations are produced beyond 3.6AU (for disk conditions at t = 5$\times$10$^{5}$ years), we can assume that the D/H
ratio will be less than 1.4$\times$10$^{-4}$. Thus it is expected that the D/H ratios for the simulated planets will be less than the currently
observed planetary values. However, given the large degree of processing this material is expected to experience both during and after planet
formation, it is likely that the planetary D/H values will increase as H is preferentially lost from the system.

\subsection{Volatile Loss\label{loss}}
The loss of volatile material (presumably through vapor loss) in large impact events can potentially be significant. \cite{okeefe} found that
for impacts between two anorthosite bodies the volume of vapor produced was less than that of the projectile for impact speeds of 15 km/s and
increased up to 10 times that of the projectile at speeds of 45 km/s. This increased up to 40 times the volume of the projectile for a collision
between an iron body and anorthosite body at 45 km/s \citep{okeefe}. Production of this vapor plume can result in significant devolatization of
the target body. For example, 20 to 60\% of water is lost from a water-bearing target body during an impact \citep{marty} while impacts into
serpentine samples produced more than 40\% devolatization for peak shock pressures above $\sim$15 GPa \citep{tyburczy2}.Similarly, shock heating
studies of Murchison meteorite samples have shown losses of up to 76\% of H from the target \citep{tyburczy}.

Although these losses of volatile material are important, detailed studies of such a loss throughout the duration of the terrestrial planet
formation process have not been undertaken. Affects such impacts into partially molten bodies, loss of elements an species other than H and
water, effective retention of the vapor plume by the growing body and subsequent re-condensation may become important during planetary
accretion.  As such, we are limited to making only first-order approximations of the amount of each element lost from the final planet due to
impacts. To do this, we need to determine both the amount of the final body that is molten and/or vaporized after each impact and the amount of
each element that would be lost from the molten phase. Using equation 9 from \cite{tonks}, the volume of melt produced by the initial shockwave
of an impact, V$_{m}$, for a hemispheric melt model is given by:
\begin{equation}
{{\rm V}_{m}} = {\rm V}_{proj}\frac{\rho_{\rm P}}{\rho_{\rm t}{\rm cos}^{1/2}i}\left(\frac{v_{i}}{v_{i}^{m}}\right)^{3/2}
\end{equation}\\

\noindent where V$_{proj}$ is the volume of the projectile, $\rho\rm _{P}$ is the density of the projectile, $\rho_{\rm t}$ is the density of
the target, $\emph{i}$ is the impact angle (as measured from the vertical), $\upsilon_{i}$ is the impact speed and $\upsilon_{i}^{m}$ is the
minimum impact speed needed to produce melting.

Thus it can be seen that to first order:

\begin{equation}
{\rm V_{m}} \propto \frac{{\rm V}_{proj} \rho_{\rm P}}{\rho_{\rm t}}v_{i}^{3/2}
\end{equation}\\

Since the fraction of the planet that is molten ($f_{m}$) is also directly proportional to the volume of the planet that is molten, we can
determine the melt fraction produced by each impact by scaling the fraction given in \cite{tonks} for a $\upsilon_{i}$ = 15 kms$^{-1}$ impact
with the v$_{i}$, M$_{proj}$ and M$_{target}$ values determined by the \cite{dave} models (assuming a uniform density for both the target and
the projectile). Although a 10-30\% difference in the melt volume produced by the truncated sphere and hemisphere melt models was identified by
\cite{tonks}, the scaling applied here is acceptable for the goals of the current study. Note that this approach assumes constant values for
$\emph{i}$ and $\upsilon_{i}^{m}$. Although these values will undoubtedly change, we are unable to provide any limitations for them at this
time. As such, we adopt the values from \cite{tonks}.

The amount of each element assumed to be lost from the molten phase is based on the condensation temperature of the element. The most volatile
element present in solid form (C, primarily present as CH$_{4}$.7H$_{2}$O in the Solar nebula) was set to lose 100\% of its mass from the melt.
For all other elements, the percentage lost was determined from the observed elemental depletions in ordinary chondrites (as compared to CI
chondrites) with respect to the 50\% condensation temperature as taken from \cite{davis}. CI chondrites are widely believed to be the most
chemically primitive meteorites. Thus the elemental depletions relative to CI chondrites observed in other meteorite classes are believed to be
due to processing of material within the solar protoplanetary disk and as such serves as an excellent proxy for the loss of volatile material.
The percentage of each element assumed to be lost in each impact event is shown in Table \ref{impacts}. Note that the condensation temperature
of O is taken to be the condensation temperature of silicate, the dominant form of O throughout the majority of the disk. The amount of each
element lost was then determined after each impact event and all lost volatile material was assumed to be permanently removed from the planet
into the surrounding disk.

This approach is obviously based on several broad assumptions and can only provide order of magnitude approximations for the loss of volatile
material. In addition to assuming details about each impact event (such as cos$^{1/2}$$\emph{i}$), this approach assumes that the target body
has completely cooled and resolidified between impacts (i.e. each impact occurs with two cold solid bodies). A hotter body produces a larger
melt fraction \citep{tonks} and is expected to result in greater loss of material from each impact event. The current simulations are based on
the equation of state for dunite obtained from the AENOS package \citep[see][for Hugoniot details]{tonks}. Variations in this may also alter the
degree of melting experienced by a body. Furthermore, we currently only consider melting produced by the initial shockwave as it moves through
the body. Detailed hydrocode simulations need to be undertaken in order to examine how volatile losses would vary under more realistic
conditions including differentiation (or lack thereof) of the target body, atmospheric losses, impacts into a still-molten embryo and
recapturing of ejected material. Finally, the current dynamical models assume perfect accretion for each of the impact events. To obtain more
realistic simulation conditions, loss of volatile material needs to be incorporated into the simulated impact events. However, this approach is
currently computationally very demanding and is not feasible.

\begin{figure}
\begin{center}
\includegraphics[width=120mm]{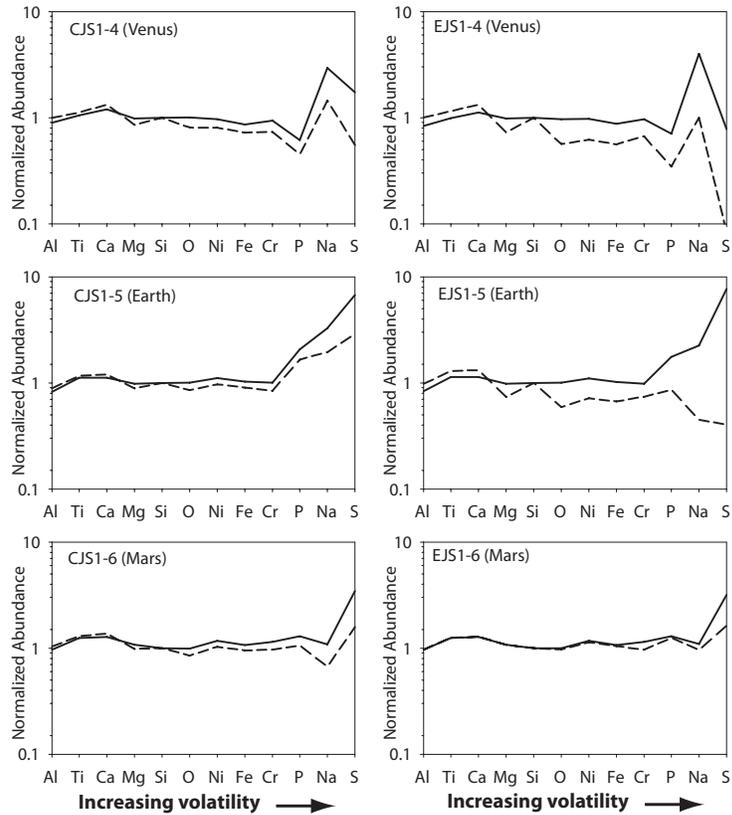} \caption[Normalized abundances for the CJS1 and EJS1 simulations after material loss
in impact events has been incorporated]{Normalized abundances for CJS1 and EJS1 simulated planets showing abundances after volatile loss was
considered. The solid line indicates the normalized abundances before volatile loss during impacts was considered while the dashed line
indicates the normalized abundance once volatile loss during impacts has been incorporated. All abundances were determined for disk conditions
at t = 5$\times$10$^{5}$ years. The terrestrial planet each simulation is normalized to is shown in parentheses. Reference Solar System
planetary abundances were taken from \cite{manda} (Venus), \cite{kandl}(Earth) and \cite{landf} (Mars). \emph{Left:} CJS1 terrestrial planets.
\emph{Right:} EJS1 terrestrial planets. \label{vol_spider} \label{lastfig}}
\end{center}
\end{figure}

As expected, loss of material through impacts reduced the amount of volatile species present in the final planetary bodies but did not
significantly alter the abundance of more refractory species. This loss of volatile elements can best be seen in the planet normalized
abundances shown in Figure \ref{vol_spider}. Clear reductions in the amount of volatile elements (specifically Na and S) can be seen, while
negligible changes are produced in the abundances of the more refractory elements (such as Fe, Cr and Mg). These reductions in the abundance of
the most volatile elements are unable to produce final planetary abundances in exact agreement with the observed planetary abundances within the
Solar System but they do represent a substantial improvement, and indicate that impact-induced melting and vaporization (and the associated loss
of material) is an important factor in determining the bulk elemental abundances of volatile species within a planet.

The average fraction of the planet melted in each individual impact event is $<$ 5\% for both simulations (2.8\% for the CJS simulations, 3.5\%
for the EJS simulations). Only one simulation (EJS1) produced an impact event large enough to melt and/or vaporize the entire target body. This
event was the 10th impact occurring on this body and thus occurred early in the dynamical formation simulation (t = 9.72$\times$10$^{6}$ years
for the dynamical simulation). As such, it is likely that although the entire body would have been disrupted by the event, a solid body may have
reformed from the remaining material afterwards. More detailed simulations are required to determine the full effects of large-scale impacts
such as this one on the terrestrial planet formation process.

\subsection{Solar Pollution}
Insignificant amounts of pollution of the Solar photosphere occurred during the terrestrial planet formation simulations. A maximum of
0.135M$_{\bigoplus}$ of solid material was added to the Sun in the CJS simulations, while the EJS simulations contributed a maximum of
1.11M$_{\bigoplus}$. The ensemble-averaged resulting solar abundances are shown in Table \ref{pollution}. The addition of solid material
generated in the CJS simulations produced no observable enrichment in the Solar spectrum. The EJS simulations, however, did produce a minor
enrichment of up to 0.02 dex for Ti and Al, 0.01 dex for C, N, Na, Mg, Al, Si, P, S, Ca, Cr, Fe and Ni and no enrichment in O. Nonetheless, this
enrichment is clearly not large enough to be definitively detected with current spectroscopic studies as many such studies produce stellar
abundances with errors equivalent to or larger than the expected enrichment (e.g. $\pm$0.03 for \cite{fv}). As such, any Solar pollution
produced by terrestrial planet formation with the Solar System is believed to be negligible.

\section{Discussion\label{sec:discussion}}
The simulations successfully produce terrestrial planets that are in excellent agreement with the terrestrial planets of the Solar System, in
terms of both their dynamics and their bulk elemental abundances. Although current simulations are unable to capture the finer details of
planetary formation (such as the composition, and sometimes the amount, of the late veneer material), the success of these simulations on a
broader scale provides us with increased confidence in the dynamical models of \cite{dave}, as well as other comparably high-resolution N-body
simulations such as \citet{raymond:2006,raymond:2007,raymond:2009} (earlier, lower-resolution simulations such as \citet{chambers2},
\citet{chambers3}, and \citet{raymond04} would likely also provide a reasonable fit in terms of bulk chemistry, but are unable to fully match
properties of the terrestrial planets such as their masses and number, formation timescale, low dynamical excitation, etc). Furthermore, it also
serves to validate the approach utilized here to combine detailed dynamical and chemical modeling together. This will allow for reliable
application of this approach not only to other dynamical models but also to other planetary systems \citep[e.g.][]{bond:thesis}.

The final bulk elemental compositions of the simulated terrestrial planets for rock forming elements are not strongly influenced by the orbital
properties or evolution of Jupiter and Saturn as can be seen by the strong similarities between the CJS and EJS simulations. The only
significant differences occur in the amount of water rich material accreted onto the final planets and the amount of late veneer material
delivered after the last major impact. This suggests that the bulk chemical evolution of the terrestrial planets is to a large extent
independent of the evolution of the giant planets. This conclusion, however, is only valid for late stage and \textit{in situ} formation after
the giant planets have formed and undertaken the vast majority of their migration. Simulations are currently running to examine the effects on
planetary composition of formation occurring during giant planet migration. As Jupiter and Saturn are not believed to have undergone extensive
migration \citep{nice1,levison,nice2}, the issue of migration is not a significant one for the Solar System. However, it will be an important
issue for extrasolar planetary systems where many are thought to have experienced large amounts of migration.

The delivery of water and other volatile species to the final terrestrial planets appears to be a normal outcome of terrestrial planet formation
within the CJS simulations, implying that it is normal for such planets to accrete significant amounts of water. As such, it negates the need
for large-scale delivery of water and other volatile species by such exotic processes as cometary impacts. The EJS simulations, on the other
hand, only produce terrestrial planets with a significant amount of water for disk conditions at times of 1$\times$10$^{6}$ years and greater.
Thus in order to produce terrestrial planets with water accreted during the formation process within the EJS configuration, time-varying
compositions need to be included within a single simulation, allowing for an increased hyrdous material content at later times. It should also
be noted that the terrestrial planets produced in the CJS simulations are likely to always accrete more water than their EJS companions.
Terrestrial planets within the CJS simulations accrete more material from the outer, more water enriched regions of the asteroid belt, thus
making them inherently water-rich compared to the terrestrial planets produced in the EJS simulations.

Later planetary processing (especially within the mantle) would have resulted in dissociation of some of the primordial water, causing a
significant amount of H to migrate to the core of the Earth or alternatively to be lost from the planet. This migration, however, would have
left a large amount of OH in the mantle, thus increasing its redox state. This process allows us to explain the evolution of the Earth's redox
state over time from the initially reduced state produced by the current simulations to the present stratified redox state. This conclusion,
however, requires a more detailed experimental understanding of the efficiency of H partioning within the mantle and core of the Earth, along
with the evolution of mantle processes and mixing.

It is interesting to note the order of magnitude difference between the CJS and EJS simulations in the amount of solid material added to the Sun
during terrestrial planet formation. This difference in accreted material is due to stronger resonances produced when Jupiter and Saturn are in
more eccentric orbits. These resonances can quickly increase the eccentricity of a body to $\sim$1, effectively driving them into the Sun and
resulting in an increased amount of accreted material. This may potentially have great bearing on extrasolar planetary host stars as many known
extrasolar planets are currently in eccentric orbits. If such an orbit always results in a greater amount of solid material being accreted by
the host star, then it may give more weight to the pollution hypothesis which has previously been suggested to explain the observed [Fe/H]
enrichment (e.g. \citealt{la,g6,murray}). However, no correlation has been found between known planetary eccentricity and the metallicity of the
host star \citep[e.g.][]{re,s2,fv,bond1}, suggesting that this is not a strong effect. We determined that a negligible amount of solid material
was added to the Solar photosphere during terrestrial planet formation and that no observable elemental enrichment would be produced. While
similar simulations are needed for extrasolar planetary systems, the current simulations support the conclusion that the observed metal
enrichment in extrasolar host stars is primordial in origin, established in the giant molecular cloud from which these systems formed as has
been concluded by other studies (such as \citealt{s1,s2,s04,s05} and \citealt{fv}). Of course, this does not rule out the possibility that
extrasolar planetary host stars may have in fact accreted a giant planet during the formation and migration process and thus display higher
levels of pollution.

Finally, biologically important elements are obviously of great interest, especially for the Earth. Of the six major biogenic elements (H, C, N,
O, S and P), four are accreted in excess by the planets during their formation (H, O, S and P). Only C and N are not accreted during planetary
formation. Both species are primarily present (under the assumption of equilibrium) as solids within the Solar nebula as clathrates (methane or
ammonia trapped in a water ice lattice) and organics. As these species only form in the outermost regions of the disk considered in the current
simulation (beyond $\sim$3.6AU for disk conditions at 5$\times$10$^{5}$ years), they could only be delivered to the final planets via cometary
impacts, migration or temporal variations within the disk or some combination of all three. Therefore delivery of material from the outer
regions of the disk are necessary in the current models for life to be able to develop. However, it is entirely possible that both C and N could
be present at low levels in solid solutions and/or liquids within the actual Solar System. Furthermore, both the CM and CR chondrites are known
to contain up to 5 wt.\% organic C. Such compositions are not captured in our current assumption of equilibrium driven condensation, indicating
the presence of another mechanism, acting to incorporate this material into carbonaceous asteroids. As such, small amounts of C and N may be
delivered during the normal accretionary process, thus reducing the requirement for external delivery of substantial amounts of material.

We did not perform a comparable chemical analysis for other recent types of terrestrial planet formation simulations, such as N-body simulations
incorporating the effects of tidal drag and sweeping secular resonances during gas disk dissipation \citep{thommes:2008}, or the hybrid
N-body/coagulation code of \citet{bromley:2006} and \citet{kenyon:2006}.  However, we expect that such simulations would also provide a
reasonable fit to the terrestrial planets in terms of bulk chemistry.  As discussed in Sec.~\ref{sec:results}, the equilibrium chemical
composition of the Solar nebula is not highly zoned, and the majority of the solid mass within the disk (at least outside of $\sim$0.75 AU) has
elemental abundances that are basically solar.  While processes that are treated in these other models [eg.~secular resonance sweeping in
\citet{thommes:2008} and gas drag on planetesimals in \citet{bromley:2006} and \citet{kenyon:2006}] may lead to a somewhat increased degree of
radial migration of material during terrestrial planet formation, that material will still have a fundamentally similar composition throughout
the disk, such that the final bulk planet chemistry is unlikely to be significantly different than what we find in this work.  More subtle
changes may occur, for example with regards to the amount of hydrated material delivered, or the amount and chemistry of the late veneer
material, which would be interesting to explore in future work.

\section{Summary}
Bulk elemental abundances have been determined for the simulated terrestrial planets of \cite{dave}. These abundances are in excellent agreement
with observed planetary values, indicating that the models of \cite{dave} are successfully producing planets comparable to those of the Solar
System in terms of both their dynamical \emph{and} chemical properties, adding greater weight to their predictive properties. Simulated redox
states are also in agreement with those predicted for the early Earth. Although differences do exist between the observed and predicted
geochemical ratios, these are believed to be a result of our assumption of equilibrium controlled compositions. Additionally, the current
simulations are unable to successfully reproduce the accretion of a late veneer of material by the early Earth, in terms of the chemistry and,
in the case of the EJS simulations, also the amount.

Significant amounts of water are accreted in the present CJS simulations, implying that the terrestrial planets in these simulations form
``wet'' and do not need significant water delivery from other sources. Terrestrial planets produced in the EJS simulations only accrete
significant amounts of water for disk conditions at 1$\times$10$^{6}$ years and beyond. N and C, however, do still need to be delivered to an
early Earth by some other process in order for life to develop.

Additionally, the bulk abundances of the major rock-forming elements for the final planets in the current simulations are not strongly dependent
on the orbital properties of the giant planets, with the CJS and EJS simulations both producing comparable results. This suggests that although
the orbits of Jupiter and Saturn are clearly of great dynamical importance to the evolution of the terrestrial planets, and can have a
significant influence in terms of water delivery, they do not exert such a large influence over the chemical evolution of the terrestrial
planets with regards to the rock-forming elements.

Finally, the pollution of the outer layers of the Sun via solid material during planetary formation produces a negligible photospheric elemental
enrichment. Assuming similar levels of pollution in other planetary systems, this in turn implies that the high metallicity trend observed in
extrasolar planetary systems is in fact primordial.

\section{Acknowledgements}
The authors thank Drew Milsom for providing the code required to calculate the disk models. We would also like to thank the anonymous reviewers
for their comments and suggestions. J. C. Bond and D. S. Lauretta were supported by grant NNX07AF96G from NASA's Cosmochemistry program. D.
O'Brien was supported by grants NNX06AC50G from NASA's Planetary Geology and Geophysics Program and NNX09AB91G from NASA's Origins of Solar
Systems Program. This paper is PSI Contribution 462.

\label{lastpage}

\bibliographystyle{elsarticle-harv}
\bibliography{references}

%% --Tables--

\clearpage
\begin{table}
\begin{center}
\caption{Properties of simulated terrestrial planets produced in the Solar System simulations of \cite{dave}. Numbering starts at 4 and increases with increasing distance from the Sun. CJS denotes the results of the circular Jupiter and Saturn simulations while EJS indicates the results of the eccentric Jupiter and Saturn simulations. \label{prop}}
\vspace{0.1in}
\begin{tabular}{|c|c|c|c|c|}
\hline {\textbf{Planet}} &
{\textbf{M}} &
{\textbf{a}} &
{\textbf{e}} &
{\textbf{i}}\\

&
&
{\textbf{(AU)}} &
&
{\textbf{($^{\circ}$)}} \\
\hline \hline

CJS1$-$4    &   1.15    &   0.63    &   0.05    &   4.36     \\
CJS1$-$5    &   0.81    &   1.21    &   0.06    &   4.85     \\
CJS1$-$6    &   0.78    &   1.69    &   0.04    &   2.06     \\
    &       &       &       &        \\
CJS2$-$4    &   0.44    &   0.55    &   0.05    &   2.61     \\
CJS2$-$5    &   0.36    &   0.69    &   0.06    &   3.54     \\
CJS2$-$6    &   1.20    &   1.10    &   0.02    &   0.38     \\
CJS2$-$7    &   0.80    &   1.88    &   0.04    &   1.83     \\
    &       &       &       &        \\
CJS3$-$4    &   0.77    &   0.62    &   0.05    &   1.54     \\
CJS3$-$5    &   1.57    &   1.14    &   0.06    &   2.11     \\
CJS3$-$6    &   0.55    &   2.09    &   0.06    &   2.07     \\
    &       &       &       &        \\
CJS4$-$4    &   1.31    &   0.66    &   0.10    &   0.60     \\
CJS4$-$5    &   1.43    &   1.54    &   0.08    &   3.64     \\
    &       &       &       &        \\
EJS1$-$4    &   0.59    &   0.56    &   0.03    &   1.69     \\
EJS1$-$5    &   0.89    &   0.84    &   0.03    &   1.24     \\
EJS1$-$6    &   0.48    &   1.29    &   0.03    &   1.55     \\
    &       &       &       &        \\
EJS2$-$4    &   0.35    &   0.50    &   0.04    &   1.61     \\
EJS2$-$5    &   0.74    &   0.75    &   0.02    &   1.07     \\
EJS2$-$6    &   0.82    &   1.17    &   0.02    &   0.71     \\
EJS2$-$7    &   0.10    &   3.19    &   0.24    &   14.95    \\
    &       &       &       &        \\
EJS3$-$4    &   0.68    &   0.58    &   0.04    &   2.12     \\
EJS3$-$5    &   0.45    &   1.52    &   0.04    &   3.20     \\
EJS3$-$6    &   0.96    &   0.96    &   0.02    &   2.10     \\
EJS3$-$7    &   0.11    &   2.08    &   0.13    &   7.83     \\
    &       &       &       &        \\
EJS4$-$4    &   0.77    &   0.68    &   0.03    &   1.48     \\
EJS4$-$5    &   0.23    &   0.48    &   0.06    &   2.29     \\
EJS4$-$6    &   0.14    &   1.07    &   0.09    &   3.82     \\
EJS4$-$7    &   0.99    &   1.33    &   0.02    &   1.59     \\

\hline
\end{tabular}
\end{center}
\end{table}

\clearpage
\begin{table}
\begin{center}
\caption[HSC Chemistry input values for Solar System Simulations]{HSC Chemistry input values for Solar System Simulations. All values are in moles. All species are assumed to initially be in their elemental and gaseous form with no other species present within the system. \label{solar-inputs}}
\vspace{0.1in}
\begin{tabular}{cc}
\hline
Element & Abundnace\\
\hline \hline
H 	&	1.00	$\times$10$^{	12	}$	 \\
He	&	8.51	$\times$10$^{	10	}$	 \\
C	&	2.45	$\times$10$^{	8	}$	 \\
N	&	6.03	$\times$10$^{	7	}$	 \\
O	&	4.57	$\times$10$^{	8	}$	 \\
Na	&	1.48	$\times$10$^{	6	}$	 \\
Mg	&	3.39	$\times$10$^{	7	}$	 \\
Al	&	2.34	$\times$10$^{	6	}$	 \\
Si	&	3.24	$\times$10$^{	7	}$	 \\
P	&	2.29	$\times$10$^{	5	}$	 \\
S	&	1.38	$\times$10$^{	7	}$	 \\
Ca	&	2.04	$\times$10$^{	6	}$	 \\
Ti	&	7.94	$\times$10$^{	4	}$	 \\
Cr	&	4.37	$\times$10$^{	5	}$	 \\
Fe	&	2.82	$\times$10$^{	7	}$	 \\
Ni	&	1.70	$\times$10$^{	6	}$	 \\
\hline
\end{tabular}
\end{center}
\end{table}

\clearpage
\begin{table}
\begin{center}
\caption[T$_{50 \% condensation}$ for the Solar System.]{T$_{50\%\:condensation}$ for a gas with Solar photosphere composition and at P =
10$^{-4}$ bar. Initial phase for each element is also listed. Values are provided from the models of \cite{lodders} for comparison.\label{cond}}
\vspace{0.3in}
\begin{tabular}{ccccc}
\hline
Element & \multicolumn{2}{c}{T$_{50\%\:condensation}$ (K)}&\multicolumn{2}{c}{Initial Phase}\\
& This Study & Lodders (2003)&Species&Formula\\
\hline \hline
Al  &   1639    &   1665    &Hibonite & CaAl$_{12}$O$_{19}$\\
C   &   $<$150  &   40  & Methane Clathrate &CH$_{4}$.7H$_{2}$O\\
Ca  &   1527    &   1505 &Hibonite & CaAl$_{12}$O$_{19}$   \\
Cr  &   1301    &   1291  & Metallic Chromium & Cr  \\
Fe  &   1339    &   1328    & Metallic Iron & Fe\\
Mg  &   1339    &   1327   & Spinel & MgAl$_{2}$O$_{4}$\\
Na  &   941 &   953 & Albite & NaAlSi$_{3}$O$_{8}$\\
Ni  &   1351    &   1348    & Metallic Nickel & Ni\\
O   &   180 &   179 &Hibonite & CaAl$_{12}$O$_{19}$\\
P   &   1309    &   1226    & Schreibersite & Fe$_{3}$P\\
S   &   658 &   655 & Troilite & FeS\\
Si  &   1329    &   1302  &  Gehlenite & Ca$_{2}$Al$_{2}$SiO$_{7}$\\
Ti  &   1580    &   1573    &Perovskite & CaTiO$_{3}$\\
\hline
\end{tabular}
\end{center}
\end{table}

\clearpage
{
\renewcommand{\baselinestretch}{1}
\small
\begin{landscape}
\begin{center}
\footnotesize
\begin{longtable}{|c|c|c|c|c|c|c|c|c|c|c|c|c|c|c|c|c|}
\caption[Predicted bulk planetary abundances for the terrestrial planets of the \cite{dave} simulations.]{Predicted bulk planetary abundances for the terrestrial planets of the \cite{dave} simulations for disk conditions at t = 5$\times$10$^{5}$yr. All values are wt\% of the final planet. CJS denotes the simulations of \cite{dave} with Jupiter and Saturn in the circular orbits predicted by the Nice model while EJS denotes the results of the simulations with Jupiter and Saturn in their current elliptical orbits. Planetary numbers start at 4 and increase with increasing distance from the Sun.} \label{5E5} \\

\hline \multicolumn{1}{|c|}{\textbf{Simulation}} &
\multicolumn{1}{c|}{\textbf{H}} &
\multicolumn{1}{c|}{\textbf{Mg}} &
\multicolumn{1}{c|}{\textbf{O}} &
\multicolumn{1}{c|}{\textbf{S}} &
\multicolumn{1}{c|}{\textbf{Fe}}&
\multicolumn{1}{c|}{\textbf{Al}}&
\multicolumn{1}{c|}{\textbf{Ca}}&
\multicolumn{1}{c|}{\textbf{Na}}&
\multicolumn{1}{c|}{\textbf{Ni}}&
\multicolumn{1}{c|}{\textbf{Cr}}&
\multicolumn{1}{c|}{\textbf{P}}&
\multicolumn{1}{c|}{\textbf{Ti}}&
\multicolumn{1}{c|}{\textbf{Si}}&
\multicolumn{1}{c|}{\textbf{N}}&
\multicolumn{1}{c|}{\textbf{C}}&
\multicolumn{1}{c|}{\textbf{Mg/Si}}\\
&
\multicolumn{1}{c|}{\textbf{wt\%}} &
\multicolumn{1}{c|}{\textbf{wt\%}} &
\multicolumn{1}{c|}{\textbf{wt\%}} &
\multicolumn{1}{c|}{\textbf{wt\%}} &
\multicolumn{1}{c|}{\textbf{wt\%}}&
\multicolumn{1}{c|}{\textbf{wt\%}}&
\multicolumn{1}{c|}{\textbf{wt\%}}&
\multicolumn{1}{c|}{\textbf{wt\%}}&
\multicolumn{1}{c|}{\textbf{wt\%}}&
\multicolumn{1}{c|}{\textbf{wt\%}}&
\multicolumn{1}{c|}{\textbf{wt\%}}&
\multicolumn{1}{c|}{\textbf{wt\%}}&
\multicolumn{1}{c|}{\textbf{wt\%}}&
\multicolumn{1}{c|}{\textbf{wt\%}}&
\multicolumn{1}{c|}{\textbf{wt\%}}&
\\\hline
\endfirsthead

\multicolumn{17}{c}%
{{\tablename\ \thetable{} -- continued from previous page}} \\
\hline \multicolumn{1}{|c|}{\textbf{Simulation}} &
\multicolumn{1}{c|}{\textbf{H}} &
\multicolumn{1}{c|}{\textbf{Mg}} &
\multicolumn{1}{c|}{\textbf{O}} &
\multicolumn{1}{c|}{\textbf{S}} &
\multicolumn{1}{c|}{\textbf{Fe}}&
\multicolumn{1}{c|}{\textbf{Al}}&
\multicolumn{1}{c|}{\textbf{Ca}}&
\multicolumn{1}{c|}{\textbf{Na}}&
\multicolumn{1}{c|}{\textbf{Ni}}&
\multicolumn{1}{c|}{\textbf{Cr}}&
\multicolumn{1}{c|}{\textbf{P}}&
\multicolumn{1}{c|}{\textbf{Ti}}&
\multicolumn{1}{c|}{\textbf{Si}}&
\multicolumn{1}{c|}{\textbf{N}}&
\multicolumn{1}{c|}{\textbf{C}}&
\multicolumn{1}{c|}{\textbf{Mg/Si}}\\
&
\multicolumn{1}{c|}{\textbf{wt\%}} &
\multicolumn{1}{c|}{\textbf{wt\%}} &
\multicolumn{1}{c|}{\textbf{wt\%}} &
\multicolumn{1}{c|}{\textbf{wt\%}} &
\multicolumn{1}{c|}{\textbf{wt\%}}&
\multicolumn{1}{c|}{\textbf{wt\%}}&
\multicolumn{1}{c|}{\textbf{wt\%}}&
\multicolumn{1}{c|}{\textbf{wt\%}}&
\multicolumn{1}{c|}{\textbf{wt\%}}&
\multicolumn{1}{c|}{\textbf{wt\%}}&
\multicolumn{1}{c|}{\textbf{wt\%}}&
\multicolumn{1}{c|}{\textbf{wt\%}}&
\multicolumn{1}{c|}{\textbf{wt\%}}&
\multicolumn{1}{c|}{\textbf{wt\%}}&
\multicolumn{1}{c|}{\textbf{wt\%}}&
\\ \hline
\endhead

\hline \multicolumn{17}{c}{{Continued on next page}} \\
\endfoot

\hline \hline
\endlastfoot

\multicolumn{17}{|c|}{t=5$\times$10$^{5}$ years}\\
\hline
CJS1-4  &   0.01    &   14.68   &   32.20   &   2.90    &   27.73   &   1.37    &   1.99    &   0.42    &   1.76    &   0.39    &   0.12    &   0.09    &   16.34   &   0.00    &   0.00    &   0.90     \\
CJS1-5  &   0.01    &   14.86   &   31.36   &   2.97    &   28.42   &   1.19    &   1.74    &   0.60    &   1.80    &   0.41    &   0.14    &   0.08    &   16.44   &   0.00    &   0.00    &   0.90     \\
CJS1-6  &   0.01    &   14.09   &   30.86   &   7.03    &   26.94   &   1.08    &   1.57    &   0.58    &   1.71    &   0.39    &   0.13    &   0.08    &   15.56   &   0.00    &   0.00    &   0.91     \\
    &       &       &       &       &       &       &       &       &       &       &       &       &       &       &       &        \\
CJS2-4  &   0.00    &   15.06   &   32.28   &   1.11    &   28.07   &   1.65    &   2.39    &   0.20    &   1.79    &   0.39    &   0.09    &   0.11    &   16.88   &   0.00    &   0.00    &   0.89     \\
CJS2-5  &   0.00    &   14.74   &   31.06   &   3.97    &   28.19   &   1.13    &   1.65    &   0.59    &   1.79    &   0.41    &   0.14    &   0.08    &   16.28   &   0.00    &   0.00    &   0.91     \\
CJS2-6  &   0.00    &   14.61   &   31.53   &   4.07    &   27.90   &   1.14    &   1.66    &   0.57    &   1.77    &   0.40    &   0.13    &   0.08    &   16.15   &   0.00    &   0.00    &   0.90     \\
CJS2-7  &   0.08    &   13.83   &   31.71   &   7.28    &   26.42   &   1.06    &   1.54    &   0.57    &   1.67    &   0.38    &   0.13    &   0.07    &   15.27   &   0.00    &   0.00    &   0.91     \\
    &       &       &       &       &       &       &       &       &       &       &       &       &       &       &       &        \\
CJS3-4  &   0.00    &   14.87   &   32.26   &   2.12    &   27.92   &   1.45    &   2.11    &   0.34    &   1.78    &   0.39    &   0.11    &   0.10    &   16.57   &   0.00    &   0.00    &   0.90     \\
CJS3-5  &   0.00    &   14.48   &   31.21   &   4.96    &   27.70   &   1.11    &   1.62    &   0.59    &   1.75    &   0.40    &   0.13    &   0.07    &   15.99   &   0.00    &   0.00    &   0.91     \\
CJS3-6  &   0.12    &   13.70   &   32.35   &   7.20    &   26.16   &   1.05    &   1.53    &   0.56    &   1.66    &   0.38    &   0.13    &   0.07    &   15.12   &   0.00    &   0.00    &   0.91     \\
    &       &       &       &       &       &       &       &       &       &       &       &       &       &       &       &        \\
CJS4-4  &   0.00    &   14.86   &   31.72   &   2.61    &   28.18   &   1.34    &   1.95    &   0.44    &   1.79    &   0.40    &   0.12    &   0.09    &   16.51   &   0.00    &   0.00    &   0.90     \\
CJS4-5  &   0.05    &   14.01   &   31.47   &   6.73    &   26.78   &   1.07    &   1.56    &   0.58    &   1.70    &   0.39    &   0.13    &   0.07    &   15.47   &   0.00    &   0.00    &   0.91     \\
    &       &       &       &       &       &       &       &       &       &       &       &       &       &       &       &        \\
%\pagebreak
EJS1-4  &   0.00    &   15.11   &   31.73   &   1.34    &   28.82   &   1.31    &   1.90    &   0.59    &   1.82    &   0.42    &   0.14    &   0.09    &   16.76   &   0.00    &   0.00    &   0.90     \\
EJS1-5  &   0.00    &   14.84   &   31.34   &   3.40    &   28.22   &   1.21    &   1.77    &   0.41    &   1.79    &   0.40    &   0.12    &   0.08    &   16.43   &   0.00    &   0.00    &   0.90     \\
EJS1-6  &   0.00    &   14.13   &   31.27   &   6.47    &   27.00   &   1.08    &   1.58    &   0.58    &   1.71    &   0.39    &   0.13    &   0.08    &   15.61   &   0.00    &   0.00    &   0.91     \\
    &       &       &       &       &       &       &       &       &       &       &       &       &       &       &       &        \\
EJS2-4  &   0.00    &   14.64   &   31.61   &   2.84    &   27.48   &   1.75    &   2.55    &   0.33    &   1.75    &   0.38    &   0.09    &   0.12    &   16.49   &   0.00    &   0.00    &   0.89     \\
EJS2-5  &   0.00    &   15.03   &   31.62   &   2.12    &   28.56   &   1.27    &   1.85    &   0.46    &   1.81    &   0.41    &   0.13    &   0.09    &   16.66   &   0.00    &   0.00    &   0.90     \\
EJS2-6  &   0.00    &   14.70   &   30.66   &   4.55    &   28.12   &   1.13    &   1.64    &   0.60    &   1.78    &   0.41    &   0.14    &   0.08    &   16.23   &   0.00    &   0.00    &   0.91     \\
EJS2-7  &   0.00    &   13.72   &   32.22   &   7.36    &   26.19   &   1.05    &   1.53    &   0.57    &   1.66    &   0.38    &   0.13    &   0.07    &   15.14   &   0.00    &   0.00    &   0.91     \\
    &       &       &       &       &       &       &       &       &       &       &       &       &       &       &       &        \\
EJS3-4  &   0.00    &   15.08   &   32.01   &   1.11    &   28.36   &   1.55    &   2.24    &   0.40    &   1.80    &   0.40    &   0.11    &   0.10    &   16.85   &   0.00    &   0.00    &   0.90     \\
EJS3-5  &   0.00    &   15.03   &   31.30   &   2.45    &   28.74   &   1.15    &   1.68    &   0.61    &   1.82    &   0.41    &   0.14    &   0.08    &   16.61   &   0.00    &   0.00    &   0.91     \\
EJS3-6  &   0.00    &   14.43   &   30.65   &   5.79    &   27.61   &   1.11    &   1.61    &   0.52    &   1.75    &   0.40    &   0.13    &   0.08    &   15.94   &   0.00    &   0.00    &   0.91     \\
EJS3-7  &   1.02    &   15.25   &   40.93   &   0.95    &   18.26   &   2.44    &   1.57    &   0.62    &   0.04    &   0.19    &   0.09    &   3.60    &   14.72   &   0.00    &   0.35    &   1.04     \\
    &       &       &       &       &       &       &       &       &       &       &       &       &       &       &       &        \\
EJS4-4  &   0.00    &   15.20   &   31.99   &   0.91    &   28.66   &   1.45    &   2.10    &   0.35    &   1.82    &   0.41    &   0.11    &   0.10    &   16.93   &   0.00    &   0.00    &   0.90     \\
EJS4-5  &   0.00    &   15.09   &   31.94   &   1.01    &   28.58   &   1.45    &   2.10    &   0.57    &   1.81    &   0.41    &   0.13    &   0.10    &   16.81   &   0.00    &   0.00    &   0.90     \\
EJS4-6  &   0.00    &   14.36   &   30.02   &   6.64    &   27.49   &   1.10    &   1.60    &   0.59    &   1.74    &   0.40    &   0.13    &   0.08    &   15.87   &   0.00    &   0.00    &   0.91     \\
EJS4-7  &   0.00    &   14.31   &   31.04   &   5.89    &   27.36   &   1.10    &   1.60    &   0.58    &   1.73    &   0.40    &   0.13    &   0.08    &   15.80   &   0.00    &   0.00    &   0.91     \\
\hline

\end{longtable}
\end{center}
\end{landscape}
}

\clearpage
\begin{table}
\begin{center}
\caption[Percentage of each element assumed to be lost in impact events.]{The percentage of each element assumed to be lost from the melt
produced by individual impact events. Values are based on depletions observed in meteorite samples and are taken from
\cite{davis}.\label{impacts}} \vspace{0.3in}
\begin{tabular}{ccc}
\hline
Element & T$_{cond}$ (K)& \% Lost\\
\hline
C   &   78  &   100.00   \\
N   &   131 &   97.69    \\
H   &   182 &   92.90    \\
S   &   704 &   49.90    \\
Na  &   958 &   32.92    \\
P   &   1248    &   16.69    \\
Cr  &   1296    &   14.32    \\
O   &   1316    &   13.37    \\
Ni  &   1353    &   11.64    \\
Fe  &   1357    &   11.45    \\
Mg  &   1397    &   9.65     \\
Si  &   1529    &   4.16     \\
Ti  &   1593    &   1.75     \\
Ca  &   1659    &   0.00     \\
Al  &   1677    &   0.00     \\

\hline
\end{tabular}
\end{center}
\end{table}

\clearpage
\begin{table}
\begin{center}
\caption[Mean change in solar photospheric abundances produced by pollution via accretion of solid material.]{Mean change in solar photospheric
abundances produced by pollution via accretion of solid material during terrestrial planet formation. All solid material migrating to within
0.1AU from the Sun during the simulations of \cite{dave} is assumed to be accreted. Change is defined as Abundance$_{\rm
after\:planet\:formation}$ - Abundance$_{\rm before\:planet\:formation}$. \label{pollution} \label{lasttable}} \vspace{0.3in}
\begin{tabular}{ccccccccc}
\hline

Element &   \multicolumn{8}{c}{Simulation}                                                           \\
    &   CJS-1   &   CJS-2   &   CJS-3   &   CJS-4   &   EJS-1   &   EJS-2   &   EJS-3   &   EJS-4    \\
\hline
Mg      &   0.00    &   0.00    &   0.00    &   0.00    &   0.01    &   0.01    &   0.01    &   0.01     \\
O   &   0.00    &   0.00    &   0.00    &   0.00    &   0.00    &   0.00    &   0.00    &   0.00     \\
S   &   0.00    &   0.00    &   0.00    &   0.00    &   0.01    &   0.01    &   0.01    &   0.01     \\
Fe      &   0.00    &   0.00    &   0.00    &   0.00    &   0.00    &   0.01    &   0.01    &   0.01     \\
Al      &   0.00    &   0.00    &   0.00    &   0.00    &   0.01    &   0.01    &   0.01    &   0.01     \\
Ca      &   0.00    &   0.00    &   0.00    &   0.00    &   0.01    &   0.01    &   0.01    &   0.01     \\
Na      &   0.00    &   0.00    &   0.00    &   0.00    &   0.01    &   0.01    &   0.01    &   0.01     \\
Ni      &   0.00    &   0.00    &   0.00    &   0.00    &   0.01    &   0.01    &   0.01    &   0.01     \\
Cr      &   0.00    &   0.00    &   0.00    &   0.00    &   0.01    &   0.01    &   0.01    &   0.01     \\
P   &   0.00    &   0.00    &   0.00    &   0.00    &   0.01    &   0.01    &   0.01    &   0.01     \\
Ti      &   0.00    &   0.00    &   0.00    &   0.00    &   0.01    &   0.01    &   0.01    &   0.01     \\
Si      &   0.00    &   0.00    &   0.00    &   0.00    &   0.01    &   0.01    &   0.01    &   0.01     \\

\hline
\end{tabular}
\end{center}
\end{table}

\clearpage

%% --Figures-- %%

\newpage
\newpage

\newpage
\newpage
\newpage
\newpage
\clearpage
\appendix
\section{\textbf{Online Material}}
This material is to appear online only as supplemental material. Table 1 lists the species (both gaseous and solid) included in the HSC chemistry calculations. The complete assemblage of all predicted bulk elemental abundances for each of the simulated planets and for each of the seven different sets of disk conditions examined are provided in Table \ref{SSwt}. Values are provided as bulk wt\% of the final planet for each set of disk conditions. Note that planetary numbers start at 4 and increase with increasing distance from the Sun.

\begin{figure}
\begin{center}
\includegraphics[width=120mm]{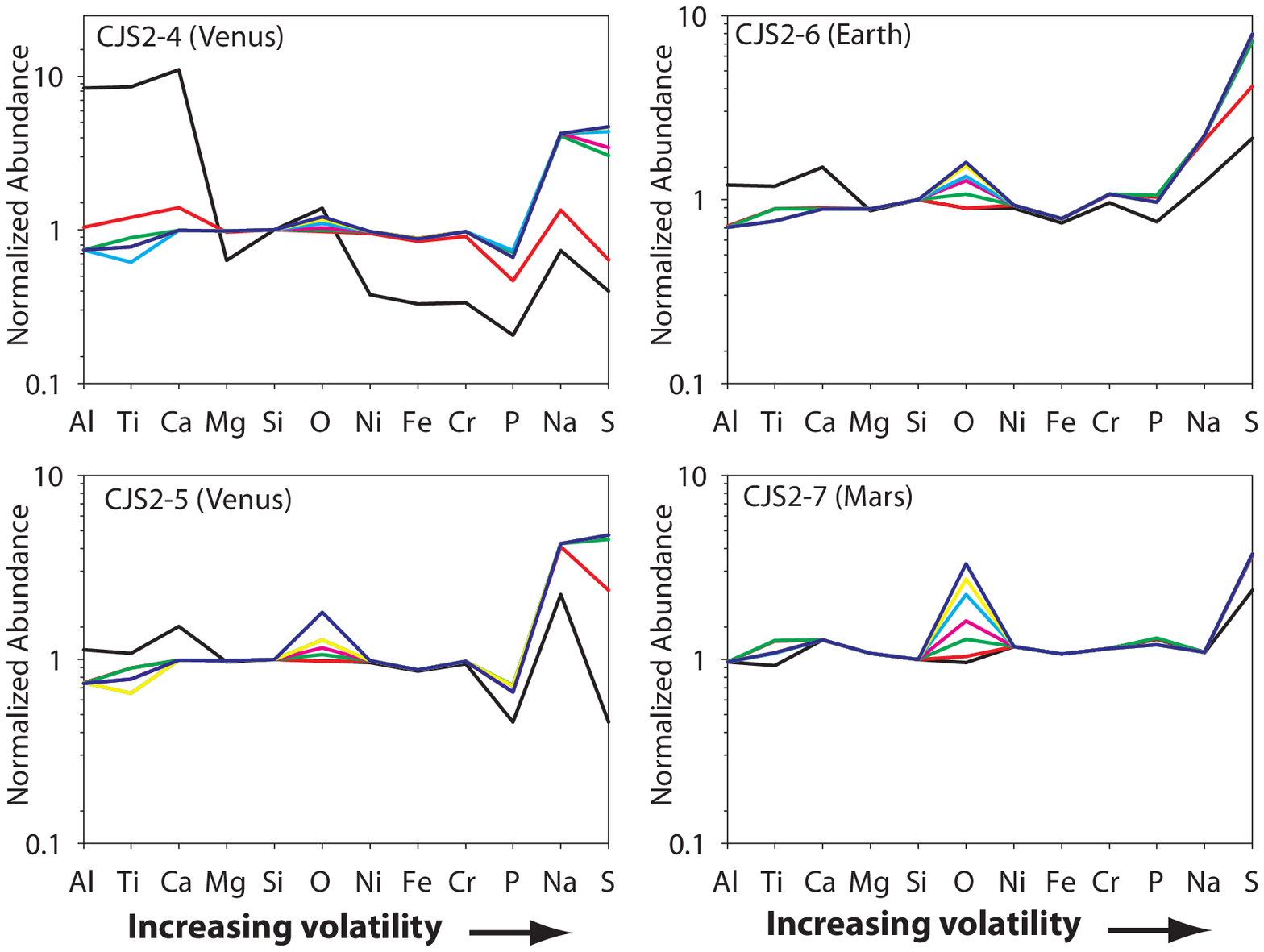} \caption[Normalized planetary abundances for CJS2 simulation.]{Normalized planetary abundances
for CJS2 simulated terrestrial planets. Values are shown for each of seven time steps considered with the following color scheme: black =
2.5$\times$10$^{5}$ years, red = 5$\times$10$^{5}$ years, green = 1$\times$10$^{6}$ years, pink = 1.5$\times$10$^{6}$ years, light blue =
2$\times$10$^{6}$ years, yellow = 2.5$\times$10$^{6}$ years and dark blue = 3$\times$10$^{6}$ years.\label{CJS2}}
\end{center}
\end{figure}

\begin{figure}
\begin{center}
\includegraphics[width=120mm]{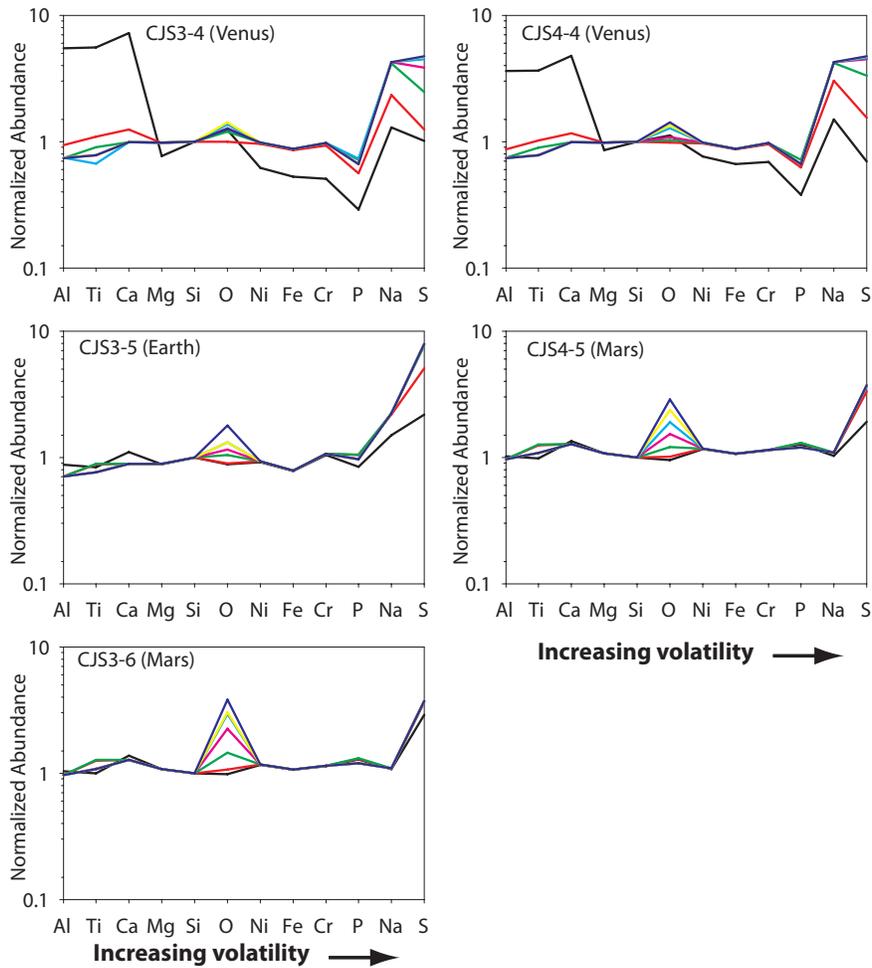} \caption[Normalized planetary abundances for CJS3 and CJS4 simulations.]{Normalized planetary abundances
for CJS3 and CJS4 simulated terrestrial planets. Values are shown for each of seven time steps considered with the following color scheme: black
= 2.5$\times$10$^{5}$ years, red = 5$\times$10$^{5}$ years, green = 1$\times$10$^{6}$ years, pink = 1.5$\times$10$^{6}$ years, light blue =
2$\times$10$^{6}$ years, yellow = 2.5$\times$10$^{6}$ years and dark blue = 3$\times$10$^{6}$ years. \emph{Left:} CJS3 simulation results.
\emph{Right:} CJS4 simulation results.\label{CJS3-4}}
\end{center}
\end{figure}

\begin{figure}
\begin{center}
\includegraphics[width=120mm]{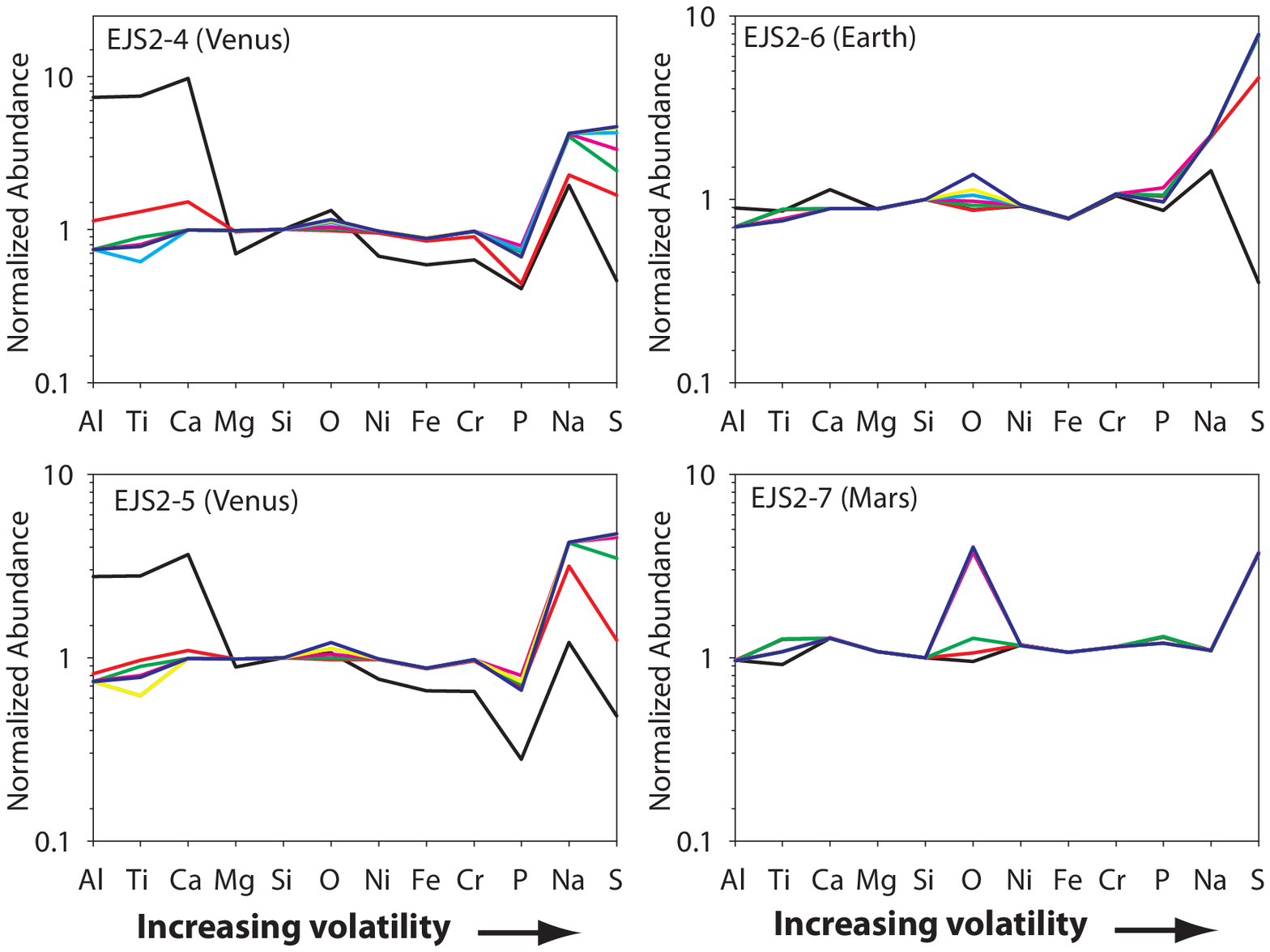} \caption[Normalized planetary abundances for EJS2 simulation.]{Normalized planetary abundances
for EJS2 simulated terrestrial planets. Values are shown for each of seven time steps considered with the following color scheme: black =
2.5$\times$10$^{5}$ years, red = 5$\times$10$^{5}$ years, green = 1$\times$10$^{6}$ years, pink = 1.5$\times$10$^{6}$ years, light blue =
2$\times$10$^{6}$ years, yellow = 2.5$\times$10$^{6}$ years and dark blue = 3$\times$10$^{6}$ years.\label{EJS2}}
\end{center}
\end{figure}

\begin{figure}
\begin{center}
\includegraphics[width=120mm]{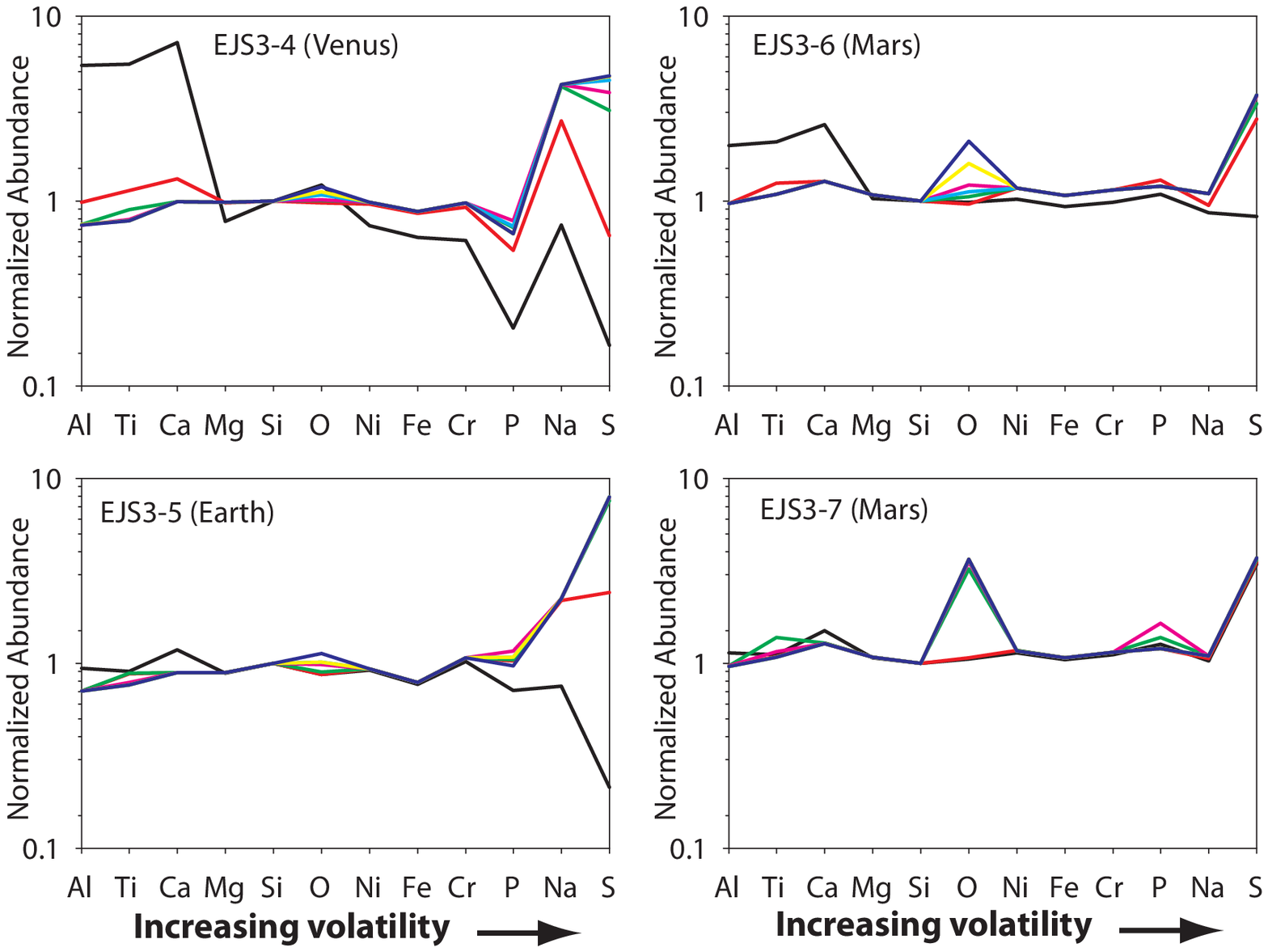} \caption[Normalized planetary abundances for EJS3 simulation.]{Normalized planetary abundances
for EJS3 simulated terrestrial planets. Values are shown for each of seven time steps considered with the following color scheme: black =
2.5$\times$10$^{5}$ years, red = 5$\times$10$^{5}$ years, green = 1$\times$10$^{6}$ years, pink = 1.5$\times$10$^{6}$ years, light blue =
2$\times$10$^{6}$ years, yellow = 2.5$\times$10$^{6}$ years and dark blue = 3$\times$10$^{6}$ years.\label{EJS3}}
\end{center}
\end{figure}

\begin{figure}
\begin{center}
\includegraphics[width=120mm]{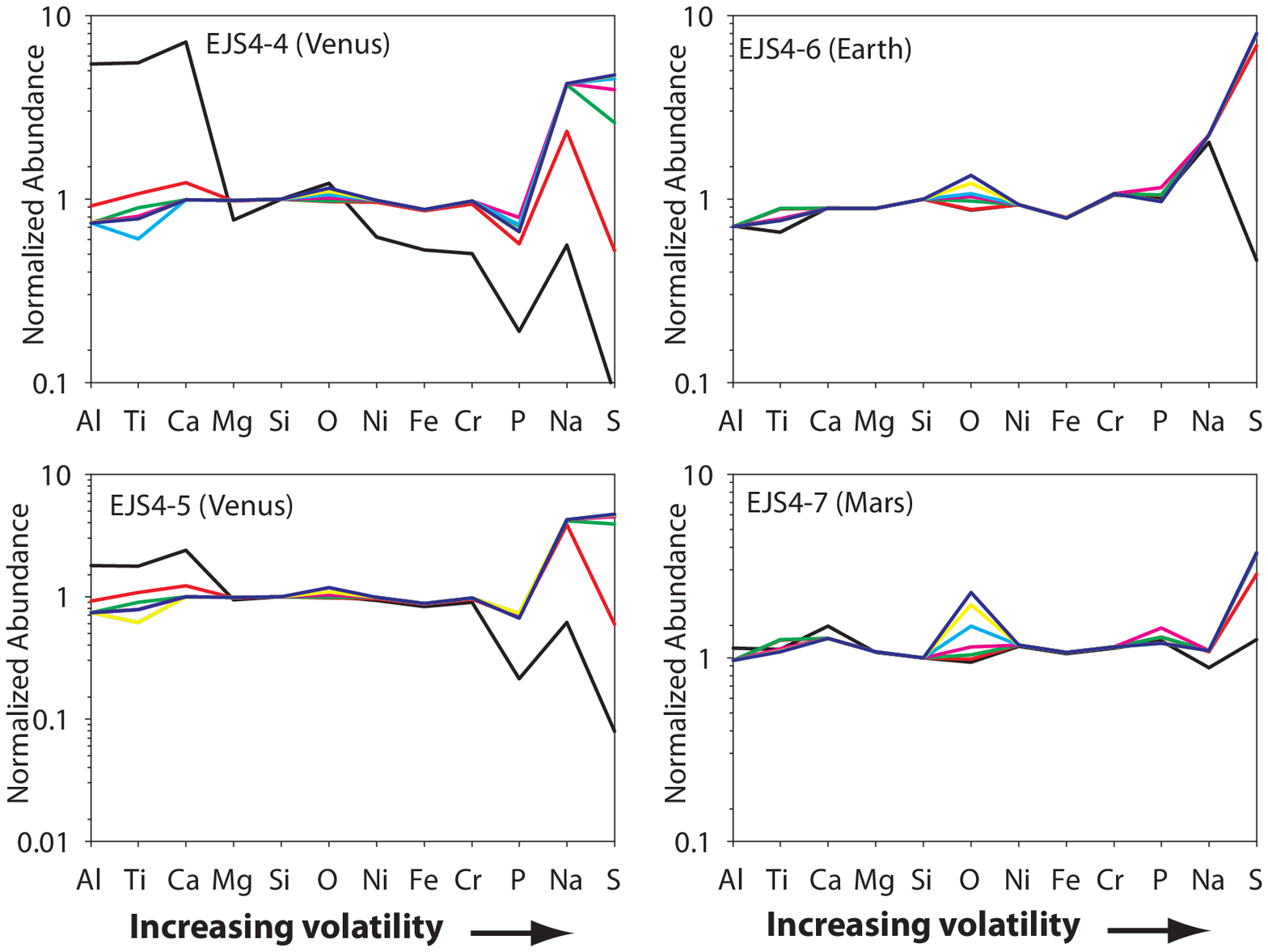} \caption[Normalized planetary abundances for EJS4 simulation.]{Normalized planetary abundances
for EJS4 simulated terrestrial planets. Values are shown for each of seven time steps considered with the following color scheme: black =
2.5$\times$10$^{5}$ years, red = 5$\times$10$^{5}$ years, green = 1$\times$10$^{6}$ years, pink = 1.5$\times$10$^{6}$ years, light blue =
2$\times$10$^{6}$ years, yellow = 2.5$\times$10$^{6}$ years and dark blue = 3$\times$10$^{6}$ years.\label{EJS4}}
\end{center}
\end{figure}

Figures \ref{CJS2} - \ref{EJS4} show the normalized abundances for each planet produced by the simulations of \cite{dave} for simulations CJS2- 4 and EJS2 - 4. Identical plots were included in the text for CJS1 and EJS1. Normalized abundances
are shown for each of the seven sets of disk conditions examined. The terrestrial planet to which each simulated planet was normalized was
determined based on the semi-major axis of the terrestrial planet and is shown in parentheses in the upper left of each plot.
\clearpage
{
\renewcommand{\baselinestretch}{1}
\small
\begin{center}
%\footnotesize
\begin{longtable}{cccc}
\caption[Predicted bulk planetary abundances for the terrestrial planets of the \cite{dave} simulations.]{Chemical species included in the equilibrium calculations of HSC
Chemistry.} \label{HSC} \\

\hline \multicolumn{4}{c}{\textbf{Gaseous Species}} \\
\hline
\endfirsthead

\multicolumn{4}{c}%
{{\tablename\ \thetable{} -- continued from previous page}} \\
\hline

\endhead

\hline \multicolumn{4}{c}{{Continued on next page}} \\
\endfoot

\hline \hline
\endlastfoot

 Al     &       CrO     &       MgOH        &       PN   \\
 AlH        &       CrOH        &       MgS     &       PO   \\
 AlO        &       CrS     &       N       &       PS   \\
 Al$_{2}$O      &       Fe      &       N$_{2}$     &       S    \\
 AlOH       &       FeH     &       NH$_{3}$        &       S$_{2}$  \\
 AlS        &       FeO     &       NO      &       SN   \\
 C      &       FeOH        &       NS      &       SO   \\
 CH$_{4}$       &       FeS     &       Na      &       SO$_{2}$ \\
\hline

\end{longtable}
\end{center}
}

\clearpage
\setlength{\LTcapwidth}{8in}
{
\renewcommand{\baselinestretch}{1}
\small
\begin{landscape}
\begin{center}
\footnotesize
\begin{longtable}{|c|c|c|c|c|c|c|c|c|c|c|c|c|c|c|c|c|}
\caption[Predicted bulk planetary abundances for the terrestrial planets of the \cite{dave} simulations.]{Predicted bulk planetary abundances for the terrestrial planets of the \cite{dave} simulations. All values are wt\% of the final planet. CJS denotes the simulations of \cite{dave} with Jupiter and Saturn in the circular orbits predicted by the Nice model while EJS denotes the results of the simulations with Jupiter and Saturn in their current elliptical orbits. Planetary numbers start at 4 and increase with increasing distance from the Sun.} \label{SSwt} \\

\hline \multicolumn{1}{|c|}{\textbf{Simulation}} &
\multicolumn{1}{c|}{\textbf{H}} &
\multicolumn{1}{c|}{\textbf{Mg}} &
\multicolumn{1}{c|}{\textbf{O}} &
\multicolumn{1}{c|}{\textbf{S}} &
\multicolumn{1}{c|}{\textbf{Fe}}&
\multicolumn{1}{c|}{\textbf{Al}}&
\multicolumn{1}{c|}{\textbf{Ca}}&
\multicolumn{1}{c|}{\textbf{Na}}&
\multicolumn{1}{c|}{\textbf{Ni}}&
\multicolumn{1}{c|}{\textbf{Cr}}&
\multicolumn{1}{c|}{\textbf{P}}&
\multicolumn{1}{c|}{\textbf{Ti}}&
\multicolumn{1}{c|}{\textbf{Si}}&
\multicolumn{1}{c|}{\textbf{N}}&
\multicolumn{1}{c|}{\textbf{C}}&
\multicolumn{1}{c|}{\textbf{Mg/Si}}\\
&
\multicolumn{1}{c|}{\textbf{wt\%}} &
\multicolumn{1}{c|}{\textbf{wt\%}} &
\multicolumn{1}{c|}{\textbf{wt\%}} &
\multicolumn{1}{c|}{\textbf{wt\%}} &
\multicolumn{1}{c|}{\textbf{wt\%}}&
\multicolumn{1}{c|}{\textbf{wt\%}}&
\multicolumn{1}{c|}{\textbf{wt\%}}&
\multicolumn{1}{c|}{\textbf{wt\%}}&
\multicolumn{1}{c|}{\textbf{wt\%}}&
\multicolumn{1}{c|}{\textbf{wt\%}}&
\multicolumn{1}{c|}{\textbf{wt\%}}&
\multicolumn{1}{c|}{\textbf{wt\%}}&
\multicolumn{1}{c|}{\textbf{wt\%}}&
\multicolumn{1}{c|}{\textbf{wt\%}}&
\multicolumn{1}{c|}{\textbf{wt\%}}&
\\\hline
\endfirsthead

\multicolumn{17}{c}%
{{\tablename\ \thetable{} -- continued from previous page}} \\
\hline \multicolumn{1}{|c|}{\textbf{Simulation}} &
\multicolumn{1}{c|}{\textbf{H}} &
\multicolumn{1}{c|}{\textbf{Mg}} &
\multicolumn{1}{c|}{\textbf{O}} &
\multicolumn{1}{c|}{\textbf{S}} &
\multicolumn{1}{c|}{\textbf{Fe}}&
\multicolumn{1}{c|}{\textbf{Al}}&
\multicolumn{1}{c|}{\textbf{Ca}}&
\multicolumn{1}{c|}{\textbf{Na}}&
\multicolumn{1}{c|}{\textbf{Ni}}&
\multicolumn{1}{c|}{\textbf{Cr}}&
\multicolumn{1}{c|}{\textbf{P}}&
\multicolumn{1}{c|}{\textbf{Ti}}&
\multicolumn{1}{c|}{\textbf{Si}}&
\multicolumn{1}{c|}{\textbf{N}}&
\multicolumn{1}{c|}{\textbf{C}}&
\multicolumn{1}{c|}{\textbf{Mg/Si}}\\
&
\multicolumn{1}{c|}{\textbf{wt\%}} &
\multicolumn{1}{c|}{\textbf{wt\%}} &
\multicolumn{1}{c|}{\textbf{wt\%}} &
\multicolumn{1}{c|}{\textbf{wt\%}} &
\multicolumn{1}{c|}{\textbf{wt\%}}&
\multicolumn{1}{c|}{\textbf{wt\%}}&
\multicolumn{1}{c|}{\textbf{wt\%}}&
\multicolumn{1}{c|}{\textbf{wt\%}}&
\multicolumn{1}{c|}{\textbf{wt\%}}&
\multicolumn{1}{c|}{\textbf{wt\%}}&
\multicolumn{1}{c|}{\textbf{wt\%}}&
\multicolumn{1}{c|}{\textbf{wt\%}}&
\multicolumn{1}{c|}{\textbf{wt\%}}&
\multicolumn{1}{c|}{\textbf{wt\%}}&
\multicolumn{1}{c|}{\textbf{wt\%}}&
\\ \hline
\endhead

\hline \multicolumn{17}{c}{{Continued on next page}} \\
\endfoot

\hline \hline
\endlastfoot

\multicolumn{17}{|c|}{t=2.5$\times$10$^{5}$ years}\\
\hline
CJS1-4  &   0.00    &   14.84   &   32.05   &   2.16    &   27.64   &   1.65    &   2.39    &   0.31    &   1.79    &   0.38    &   0.09    &   0.09    &   16.61   &   0.00    &   0.00    &   0.89     \\
CJS1-5  &   0.00    &   14.86   &   31.94   &   1.61    &   28.12   &   1.66    &   2.41    &   0.35    &   1.79    &   0.40    &   0.11    &   0.09    &   16.66   &   0.00    &   0.00    &   0.89     \\
CJS1-6  &   0.00    &   14.73   &   31.43   &   3.34    &   28.08   &   1.27    &   1.84    &   0.58    &   1.78    &   0.41    &   0.13    &   0.07    &   16.34   &   0.00    &   0.00    &   0.90     \\
    &       &       &       &       &       &       &       &       &       &       &       &       &       &       &       &        \\
CJS2-4  &   0.00    &   8.29    &   38.76   &   0.58    &   9.23    &   11.24   &   16.12   &   0.09    &   0.60    &   0.12    &   0.03    &   0.66    &   14.27   &   0.00    &   0.00    &   0.58     \\
CJS2-5  &   0.00    &   14.89   &   32.06   &   0.78    &   28.41   &   1.77    &   2.58    &   0.33    &   1.80    &   0.41    &   0.09    &   0.10    &   16.77   &   0.00    &   0.00    &   0.89     \\
CJS2-6  &   0.00    &   14.60   &   32.19   &   2.18    &   27.09   &   1.94    &   2.81    &   0.35    &   1.74    &   0.37    &   0.10    &   0.11    &   16.52   &   0.00    &   0.00    &   0.88     \\
\hline

\end{longtable}
\end{center}
\end{landscape}
}

\clearpage
\clearpage
\clearpage
\clearpage
\clearpage

\end{document}